\documentclass[useAMS,usenatbib]{mnras}
\usepackage{natbib}
\usepackage{amsmath}
\usepackage{url}
\usepackage{longtable}
\usepackage{aas_macros}
\usepackage{amssymb}
\usepackage{graphicx}
\usepackage{deluxetable}
\usepackage{pdflscape}
\usepackage{multirow}
\usepackage{color}

\newcommand{\gri}{\protect\hbox{$gri$} }

\newcommand{\grips}{\protect\hbox{$gri_{\rm P1}$}}
\newcommand{\grizps}{\protect\hbox{$griz_{\rm P1}$}}
\newcommand{\grizyps}{\protect\hbox{$grizy_{\rm P1}$}}
\newcommand{\gps}{\protect\hbox{$g_{\rm P1}$}}
\newcommand{\rps}{\protect\hbox{$r_{\rm P1}$}}
\newcommand{\ips}{\protect\hbox{$i_{\rm P1}$}}
\newcommand{\zps}{\protect\hbox{$z_{\rm P1}$}}
\newcommand{\yps}{\protect\hbox{$y_{\rm P1}$}}
\newcommand{\wps}{\protect\hbox{$w_{\rm P1}$}}

\newcommand{\jhk}{\protect\hbox{$J\!H\!K$} }

\newcommand{\about}{$\sim\!\!$~}

\def\lsim{\hbox{\rlap{\raise 0.425ex\hbox{$<$}}\lower 0.65ex\hbox{$\sim$}}}
\def\gsim{\hbox{\rlap{\raise 0.425ex\hbox{$>$}}\lower 0.65ex\hbox{$\sim$}}}

\def\arcsec{\hbox{$^{\prime\prime}$}}

\newcommand{\dof}{\rm dof}

\newcommand{\ntot}{342}
\newcommand{\nsn}{225}
\newcommand{\ncosmo}{180}
\newcommand{\nsnap}{169}
\newcommand{\ngal}{70}
\newcommand{\ngalper}{32.7}
\newcommand{\sigint}{0.111}
\newcommand{\sigchi}{0.105}
\newcommand{\fracasassnone}{40}
\newcommand{\fracasassntwo}{41}
\newcommand{\fracpsstone}{21}
\newcommand{\fracpssttwo}{45}
\newcommand{\repeatfrac}{1.1}
\newcommand{\weatherfrac}{39}
\newcommand{\medcad}{6}
\newcommand{\medseeg}{1.28}
\newcommand{\medseer}{1.16}
\newcommand{\medseei}{1.05}
\newcommand{\medseez}{1.02}
\newcommand{\medz}{0.033}
\newcommand{\medmuerr}{1.10}
\newcommand{\medx}{0.048}
\newcommand{\medc}{$-0.021$}

\title[Foundation Supernova Survey]{The Foundation Supernova Survey:
  Motivation, Design, Implementation, and First Data Release}

\def\ucsc{1}
\def\uc{2}
\def\stsci{3}
\def\rut{4}
\def\jhu{5}
\def\cfa{6}
\def\ifa{7}
\def\nd{8}
\def\moore{9}
\def\uiuc{10}
\def\qub{11}

\begin{document}

\author[Foley, Scolnic, Rest et~al.]{Ryan~J.~Foley$^{\ucsc}$\thanks{E-mail:foley@ucsc.edu}, Daniel~Scolnic$^{\uc}$, Armin~Rest$^{\stsci}$, 
S.~W.~Jha$^{\rut}$,
Y.-C.~Pan$^{\ucsc}$,
\newauthor
A.~G.~Riess$^{\jhu, \stsci}$,
P.~Challis$^{\cfa}$,
K.~C.~Chambers$^{\ifa}$,
D.~A.~Coulter$^{\ucsc}$,
K.~G.~Dettman$^{\rut}$,
\newauthor
M.~M.~Foley$^{\nd}$,
O.~D.~Fox$^{\stsci}$, 
M.~E.~Huber$^{\ifa}$,
D.~O.~Jones$^{\ucsc}$,
C.~D.~Kilpatrick$^{\ucsc}$,
\newauthor
R.~P.~Kirshner$^{\cfa, \moore}$,
A.~S.~B.~Schultz$^{\ifa}$,
M.~R.~Siebert$^{\ucsc}$,
H.~A.~Flewelling$^{\ifa}$,
B.~Gibson$^{\ifa}$,
\newauthor
E.~A.~Magnier$^{\ifa}$,
J.~A.~Miller$^{\uiuc}$,
N.~Primak$^{\ifa}$,
S.~J.~Smartt$^{\qub}$,
K.~W.~Smith$^{\qub}$
\newauthor
R.~J.~Wainscoat$^{\ifa}$,
C.~Waters$^{\ifa}$,
M.~Willman$^{\ifa}$\\
$^{\ucsc}$Department of Astronomy and Astrophysics, University of California, Santa Cruz, CA 95064, USA\\
$^{\uc}$Kavli Institute for Cosmological Physics, The University of Chicago, Chicago, IL 60637, USA\\
$^{\stsci}$Space Telescope Science Institute, 3700 San Martin Drive, Baltimore, MD 21218, USA\\
$^{\rut}$Department of Physics and Astronomy, Rutgers, The State University of New Jersey, 136 Frelinghuysen Road, Piscataway, NJ 08854, USA\\
$^{\jhu}$Department of Physics and Astronomy, Johns Hopkins University, Baltimore, MD, USA\\
$^{\cfa}$Harvard College Observatory, Harvard University, 60 Garden Street, Cambridge MA 02138, USA\\
$^{\ifa}$Institute for Astronomy, University of Hawaii, 2680 Woodlawn Drive, Honolulu, HI 96822, USA\\
$^{\nd}$Department of Physics, University of Notre Dame, Notre Dame, IN 46556, USA\\
$^{\moore}$Gordon and Betty Moore Foundation, 1661 Page Mill Road, Palo Alto, CA 94304, USA\\
$^{\uiuc}$Astronomy Department, University of Illinois at Urbana--Champaign, 1002 W.\ Green Street, Urbana, IL 61801, USA\\
$^{\qub}$Astrophysics Research Centre, School of Mathematics and Physics, Queen's University Belfast, Belfast BT7 1NN\\
}

\date{Accepted  . Received   ; in original form  }
\pagerange{\pageref{firstpage}--\pageref{lastpage}} \pubyear{2017}
\maketitle
\label{firstpage}

\begin{abstract}
  The Foundation Supernova Survey aims to provide a large,
  high-fidelity, homogeneous, and precisely-calibrated low-redshift
  Type Ia supernova (SN Ia) sample for cosmology.  The calibration of
  the current low-redshift SN sample is the largest component of
  systematic uncertainties for SN cosmology, and new data are
  necessary to make progress.  We present the motivation, survey
  design, observation strategy, implementation, and first results for
  the Foundation Supernova Survey.  We are using the Pan-STARRS
  telescope to obtain photometry for up to 800 SNe~Ia at $z \lesssim
  0.1$.  This strategy has several unique advantages: (1) the
  Pan-STARRS system is a superbly calibrated telescopic system, (2)
  Pan-STARRS has observed 3/4 of the sky in \grizyps\ making future
  template observations unnecessary, (3) we have a well-tested
  data-reduction pipeline, and (4) we have observed \about 3000
  high-redshift SNe~Ia on this system.  Here we present our initial
  sample of \nsn\ SN~Ia \grizps\ light curves, of which \ncosmo\ pass
  all criteria for inclusion in a cosmological sample.  The Foundation
  Supernova Survey already contains more cosmologically useful SNe~Ia
  than all other published low-redshift SN~Ia samples combined.  We
  expect that the systematic uncertainties for the Foundation
  Supernova Sample will be 2--3 times smaller than other low-redshift
  samples.  We find that our cosmologically useful sample has an
  intrinsic scatter of \sigint~mag, smaller than other low-redshift
  samples.  We perform detailed simulations showing that simply
  replacing the current low-redshift SN~Ia sample with an equally
  sized Foundation sample will improve the precision on the dark
  energy equation-of-state parameter by 35\%, and the dark energy
  figure-of-merit by 72\%.
\end{abstract}



\section{Introduction}\label{s:intro}

Observations of Type Ia supernovae (SNe~Ia) led to the discovery that
the Universe's expansion is currently accelerating
\citep{Riess98:Lambda, Perlmutter99}.  SNe~Ia continue to be a mature
and important cosmological tool \citep[e.g.,][]{Suzuki12, Betoule14,
  Rest14, Jones17, Scolnic17}.  Further observations of SNe~Ia will be
critical to improved understanding of the nature of dark energy,
perhaps the most puzzling open problem in all of physics.

Several observational methods have been employed to address this
problem.  This multi-probe approach has dramatically improved our
ability to constrain dark energy parameters such as its equation of
state, $w = P/\rho c^{2}$, where $P$ is its pressure and $\rho$ is its
density, and any evolution of $w$ with redshift or cosmic time.
Although in many ways the simplest explanation consistent with current
observations is that dark energy is a cosmological constant with $w =
-1$, recent results hint at a possible deviation from this
\citep[e.g.,][]{Planck14, Rest14}.  At the very least, this
demonstrates that there is currently no definitive answer to exactly
what is driving the Universe's accelerated expansion, and new data and
analyses are required to make progress.  To that end, SNe~Ia are still
an exquisite tool with which we can precisely measure the expansion
history of the Universe.

Since dark energy may evolve over cosmic time, one can describe a more
general parameterization of the equation-of-state parameter,
\begin{align}
  w(a) &= P(a) / \rho(a) c^{2}\\
       &= w_{0} + (1 - a) w_{a},
\end{align}
where $a = (1 + z)^{-1}$ is the scale factor of the Universe, $w_{0}$
is the current equation-of-state parameter of dark energy, and $w_{a}$
parameterizes the evolution of the equation-of-state parameter
\citep[see e.g.,][]{Linder03}.  In order to quantify our knowledge (or
ignorance) of dark energy, the Dark Energy Task Force (DETF) defined a
Figure of Merit (FoM) which is equal to the inverse of the area
enclosed within the 95\% confidence contour in the $w_{0}$--$w_{a}$
plane (and equivalent to the inverse of the square root of the
determinant of the covariance matrix for $w_{0}$ and $w_{a}$;
\citealt{Albrecht06}; \citealt{Wang08:fom}).  With this choice, larger
FoMs indicate increasing knowledge of dark energy.  As a point of
reference, a recent analysis combining multiple probes had a FoM of
32.6 \citep{Alam17}.

Another fundamental cosmological parameter measurable with
observations of SNe~Ia is the Hubble constant, $H_{0}$.  The Hubble
constant is also critically related to the age of the Universe and
indirectly constrains $w$ \citep{Hu05}.  Over the last decade, the
uncertainty on the Hubble constant has been significantly reduced
\citep[e.g.,][see \citealt{Freedman10} for a recent
review]{Riess09:h0, Riess11, Freedman12}.  The most recent progress in
constraining the Hubble constant has reduced the uncertainty to 2.4\%
\citep{Riess16}.  Much of this success has been the result of linking
measurements of precise, abundant, but relatively faint distance
indicators such as Cepheid variables with measurements of precise,
less frequent, and relatively luminous distance indicators such as
SNe~Ia.  Recently, direct measurements of $H_{0}$ seem in conflict (at
the 3.4--$\sigma$ level) with the inferred values derived from
indirect techniques such as combining measurements of the cosmic
microwave background with large-scale structure \citep{Planck16,
  Alam17}.  This tension could be the result of subtle systematic
uncertainties in the various measurements, but could also point to
interesting new physics such as an additional relativistic species
akin to neutrinos, but distinct from the known members of that family.

At this point, we have amassed enough SNe that the precision for $w$
is not limited by statistical precision, but rather systematic floors
in our ability to measure the distances to SNe~Ia prevent further
progress.

Currently, the largest systematic uncertainty for high-$z$ SN
cosmology is photometric calibration \citep{Conley11, Betoule14,
  Scolnic14:ps1, Scolnic17}.  Cosmological constraints from SNe~Ia are
derived by comparing the distances of low- to high-$z$ SNe, with the
low-$z$ sample providing an `anchor' to the high-$z$ sample.  At the
moment, the heterogeneous low-$z$ SN~Ia sample is a larger source of
calibration uncertainty than the high-$z$ samples, and all other
individual systematic uncertainties.  Additionally, the high-$z$
samples are now much larger than the low-$z$ sample
\citep[e.g.,][]{Scolnic17}, meaning that the statistical weight of
each low-$z$ SN is currently larger than each high-$z$ SN.  Any work
on high-$z$ samples will have a marginal effect on $w$ until we
improve the low-$z$ sample \citep{Astier14, Spergel15}.

Here we present the strategy, implementation, and first results for
the Foundation Supernova Survey.  The Foundation Supernova Survey is
designed to replace the current low-$z$ ($z < 0.1$) SN~Ia sample with
a large, homogeneous, high-fidelity sample which can be used as the
low-$z$ anchor (and {\it foundation}) for future cosmological
analyses.  This sample will address the largest uncertainty in SN
cosmology, significantly reducing the total uncertainty for $w$.  The
Foundation Supernova Survey will fulfill a large portion, and perhaps
all, of the {\it WFIRST} low-$z$ sample requirement \citep{Spergel15}.
Since the Large Synoptic Survey Telescope (LSST) will saturate at $m
\approx 17$~mag, the Foundation Supernova Survey may be the
fundamental low-$z$ SN~Ia sample, especially for measurements of
$H_{0}$, for at least a decade.

This manuscript is structured as follows: Section~\ref{s:design}
describes the overall survey requirements, strategy, and design
choices; Section~\ref{s:obs} presents observations and data reduction
for our initial sample of SNe observed in the first few months of the
survey; Section~\ref{s:results} presents our initial results; and
Section~\ref{s:disc} discusses our future goals and possible
improvements.


\section{Survey Design}\label{s:design}

\subsection{Motivation}

Current SN~Ia cosmology analyses \citep[e.g.,][]{Rest14} use \about
200 low-$z$ SNe~Ia from six different low-$z$ surveys, each on a
different photometric system with different cadences and systematic
errors.  The heroic effort of creating the current low-$z$ SN~Ia
sample was performed over more than two decades, and some of these
data were critical in the original detection of the accelerating
universe.

The \citet{Rest14} analysis, which used a particularly large low-$z$
sample \citep[larger than e.g.,][]{Betoule14}, included samples from
the following surveys: {\bf Cal\'{a}n/Tololo} \citep[16 SNe after
various quality cuts, including a redshift cut to only include SNe in
the Hubble flow ($z > 0.015$); mostly CTIO 0.9-m + Tek1, Tek2, Tek3,
Tek IV, TI2, TI3;][]{Hamuy96:lc}, {\bf CfA1} \citep[5 SNe; mostly
1.2-m FLWO + thick/thin CCDs;][]{Riess99:lc}, {\bf CfA2} \citep[19
SNe; 1.2-m + Andycam, 4Shooter;][]{Jha06:lc}, {\bf CfA3} \citep[85
SNe; 1.2-m + 4Shooter, Minicam, and Keplercam;][]{Hicken09:lc}, {\bf
  CfA4} \citep[43 SNe; 1.2-m + Keplercam;][]{Hicken12}, {\bf CSP}
\citep[45 SNe; Swope + SITe3; templates from du~Pont;][]{Contreras10},
and 8 {\bf ``other SNe''} included by \citet{Jha06:lc}.  Even without
considering multiple filter systems for a given telescope/camera, the
low-$z$ SN~Ia sample is constructed from $>$13 systems, with the
largest and second-largest homogeneous samples being the
1.2-m/Keplercam (used for part of CfA3 and CfA4, respectively) and
1.2-m/4Shooter systems with 93 and 32 SNe~Ia, respectively.  One could
also include other surveys such as {\bf LOSS} \citep[mostly KAIT +
Apogee, Apogee2, FLI;][]{Ganeshalingam10}, but that only increases the
inhomogeneity.

The cosmological systematic uncertainty associated with this
heterogeneous low-$z$ sample is perhaps best demonstrated by the shift
in the measured value of $w$ when we exclude a survey.  When
\citet{Scolnic14:ps1} excluded the CSP sample, $w$ shifted by $-0.021$
even when combining with cosmic microwave background (CMB) and baryon
acoustic oscillation (BAO) data.  On the other hand, excluding CfA1,
CfA2, and the ``other'' SNe shifted $w$ by $+0.022$.  This $>$4\%
overall shift is larger than that expected from the increased
statistical uncertainty from removing the samples and an indication of
a large systematic bias --- but it does not fully capture our
ignorance.  There remain fundamental and irreducible systematic
uncertainties associated with the data where the cameras, filters,
{\it and even telescopes} no longer exist.

Additionally, the calibration errors from the many different low-$z$
samples propagate through the analysis in multiple ways as we fit {\it
  all} SN~Ia light curves using models determined primarily from the
low-$z$ SNe.  These systematic uncertainties have only begun to be
unraveled \citep{Mosher14} and likely result in a moderate bias in our
measurement of $w$.

Because the older surveys did not always start monitoring the SN
before peak brightness, did not always have sufficient filter coverage
and/or cadence, and observed many SNe~Ia that were not in the Hubble
flow ($z > 0.015$), only 45\% of the observed low-$z$ SNe are included
in current analyses.  Many records from 20 years ago are now lost, and
therefore, we cannot precisely model the selection bias for the
current sample.  Since past low-$z$ samples are a subset of the SNe
discovered during that era, and most low-$z$ SN surveys targeted large
galaxies, there are well known selection biases \citep{Scolnic14:col}
in the colour and luminosity distributions of the sample.  These
biases also propagate into the systematic uncertainty through the SN
colour model, the host galaxy dependence, and the selection bias
itself.

We have concluded that the current heterogeneous low-$z$ SN~Ia sample
is limiting current cosmological analyses.  Although one can make some
modest improvements in the sample with great effort \citep{Scolnic15},
ultimately, the sample must be replaced to break through the current
systematic floors.  Fortunately, the SN discovery rate, particularly
for SNe before maximum brightness, has increased significantly over
the past few years \citep[e.g.,][]{Gal-Yam13}, and a sufficient number
of SNe~Ia are discovered to completely replace the number of SNe~Ia in
the low-$z$ sample within 1--2 years.

\subsection{Telescopic System Requirements}\label{ss:tel_req}

When considering replacing the low-$z$ SN sample, one must identify
the limitations of the current sample.  The photometric calibration is
clearly the largest hurdle, which can be separated into two main
components: the characterization of a given system and placing that
system on a physical scale.

If all potential calibration systematic uncertainties are properly
assessed for all surveys, one could combine all data to ``average''
any potential biases in a single survey.  However, in practice, some
uncertainties are highly correlated between systems (such as absolute
flux calibration) and some systematic uncertainties may be
underestimated \citep{Scolnic15}.  As such, one must be careful in
choosing a specific sample for inclusion in an analysis.

Although we make no specific recommendations for inclusion/exclusion
here, we note that every additional telescope, camera, and/or filter
will result in increased calibration uncertainty.  While many
different systems may reduce the overall systematic uncertainty, a
practical place to start is to calibrate a single telescopic system as
well as possible.

To do this, the telescopic system should be externally well calibrated
and self-consistent.  That is, the absolute and relative photometric
uncertainty of the system should be minimized.  One can significantly
improve the relative photometric calibration by linking all SN fields
together through overlapping observations.  This
``\"{u}bercalibration'' has been performed for the large-footprint
Sloan Digital Sky Survey (SDSS) and Pan-STARRS1 (PS1) surveys
\citep{Padmanabhan08, Schlafly12}, but is difficult to do with any
small field-of-view camera.  Moreover, multiple observations of the
sky spread out in time improves this calibration significantly.
Precise absolute calibration can be obtained by observing
spectrophotometric standard stars.

Next, the filter response functions should be precisely measured.
Ideally, this is done with a tunable laser diode or monochromator to
measure the full system throughput as a function of wavelength
\citep[e.g.,][]{Rheault10, Stubbs12}.

Finally, an ideal system should have already observed a large {\it
  high-redshift} SN~Ia sample.  Such a sample would allow a
measurement of cosmological samples using a single telescopic system,
reducing the number of free parameters in the systematic error budget
to the minimal case.

Of all available telescopic systems, only PS1 has all desired
characteristics.  It is therefore an ideal platform upon which one
could construct a new low-$z$ SN~Ia sample.

\subsection{The Pan-STARRS-1 System}

The PS1 system is a high-etendue wide-field imaging system, designed
for dedicated survey observations.  The system is installed on the
peak of Haleakala on the island of Maui in the Hawaiian island chain.
We provide below a terse summary of the PS1 survey instrumentation.  A
more complete description of the PS1 system, both hardware and
software, is provided by \citet{Kaiser10}, \citet{Chambers17}, and
references therein.

The PS1 optical design \citep{Hodapp04} uses a 1.8~meter diameter
$f$/4.4 primary mirror, and a 0.9~m secondary.  The telescope delivers
images with low distortion over a field diameter of 3.3 degrees. The
1.4~Gigapixel PS1 imager \citep{Tonry09} comprises a total of 60 $4800
\times 4800$ pixel detectors, with 10~$\mu$m pixels that subtend
0\farcs258.  The detectors are back-illuminated CCDs manufactured by
Lincoln Laboratory, which are read out using a StarGrasp CCD
controller in 7 seconds for a full unbinned image. Initial performance
assessments are presented in \citet{Onaka08}.

The PS1 observations are obtained through a set of 5 broadband
filters, which we have designated as \gps, \rps, \ips, \zps, and \yps\
(\grizyps). The PS1 system also has a wide $gri$-composite filter
\wps, which is not currently used by our survey.  Instrumental
response functions for the PS1 filters have been measured by
\citet{Tonry12}.

PS1 has unique and important characteristics for building a low-$z$
SN~Ia sample.  {\bf (1)} The telescope has already observed 3/4
(3$\pi$) of the sky in 5 filters.  {\bf (2)} PS1 has a superb relative
photometric calibration down to a few mmag \citep{Schlafly12,
  Magnier13}.  Because of its excellent calibration, PS1 has been used
to re-calibrate SDSS \citep{Finkbeiner16}.  {\bf (3)} PS1 has one of
the best measured instrument response functions, making SN photometry
extremely precise \citep{Stubbs10, Magnier17}.  {\bf (4)} There
already exists a very large PS1 high-$z$ SN~Ia sample (\about 400
spectroscopically confirmed SNe~Ia; \citealt{Rest14}, Scolnic et~al.,
in prep.).  {\bf (5)} Because of our familiarity with PS1, we can use
our robustly tested and optimized pipeline to quickly produce light
curves from our newly acquired data.  We detail these advantages
below.

\begin{enumerate}
\item {\bf Sky coverage:}
  The PS1 3$\pi$ survey was completed at the beginning of 2014 (see
  \citealt{Chambers17} for a description of the 3$\pi$ survey
  strategy).  Roughly 12 images of every position north of $-30$
  declination were observed in 5 filters (60 total images).  The
  stacked images are roughly as deep as SDSS in $gr$, but as much as
  0.5 -- 1~mag deeper in $iz$, with typically better seeing
  \citep{Metcalfe13}.  Multiple images, each providing independent
  measurements, improve the absolute calibration and reduce cosmetic
  defects such as bad pixels, chip gaps, satellite trails, etc.  The
  PS1 Science Consortium has publicly released the stacked sky images
  and catalogues through the the Space Telescope Science Institute
  (STScI), and the data can be accessed through MAST, the Mikulski
  Archive for Space Telescopes \footnote{http://panstarrs.stsci.edu/}.

  The field for any SN that PS1 can observe at relatively low airmass
  has already been observed as part of the 3$\pi$ survey.  This
  immediately provides photometric zeropoints and templates for all
  new SNe~Ia.  Without the need for templates after the SN has faded,
  the amount of telescope time is reduced.  But more importantly,
  final photometry can be measured immediately; one does not need to
  wait a year (or longer) for the SN to fade before getting templates.

\item {\bf Well-calibrated photometric system:}
  The PS1 system already has a calibration that is significantly more
  uniform than any low-$z$ survey.  In fact, the PS1 calibration is at
  least as good as SDSS, where PS1 images can now be used to correct
  SDSS photometry \citep{Finkbeiner16}.  The PS1 calibration will keep
  improving in the coming years as PS1 continues to observe and
  calibration methodology improves.  By linking to the 3$\pi$ data, SN
  photometry can be calibrated to $<$5~mmag precision.

  This is in contrast to other low-$z$ surveys, which observe Landolt
  or other spectrophotometric standard stars during photometric
  nights, and then convert those magnitudes to the standard AB system
  to determine photometric zeropoints.

\item {\bf Precise and Accurate Instrument Response:}
  An accurate instrument response function is essential for the
  spectral-model fitting needed to compare SNe~Ia at different
  redshifts.  The state-of-the art method to measure the instrument
  response function is to use a NIST-calibrated photodiode in
  combination with a tunable laser system to pass a known amount of
  light through the entire telescope system in small wavelength
  increments \citep{Stubbs07, Stubbs10, Stubbs12}.

  This is a relatively technical, time-intensive, and expensive
  measurement.  Even with infinite resources, we would be unable to
  make this measurement for the majority of the current low-$z$ sample
  --- for many, the cameras have been de-commissioned and the filters
  have degraded with time.

  A very large amount of person power, money, and telescope time has
  already been put into characterizing the PS1 system using a tunable
  laser, and as a result, it has one of the best-measured instrument
  response functions with sub-nm precision for the effective filter
  wavelengths.

  Furthermore, repeated PS1 imaging of 10 Medium-Deep fields (MDFs)
  over 4.5 years provides hundreds of measurements of thousands of
  stars to determine if the system is evolving with time.  Thus far,
  we have constrained any evolution to $<$3~mmag over the first two
  years of the PS1 survey (Scolnic et~al., in prep.).

  Although significant effort has already been put into producing the
  current PS1 calibration, it can be further improved.  One of the
  greatest advantages of the PS1 survey is that it covers such a large
  portion of the sky, including the majority of {\it HST} Calspec
  standards \citep[with a magnitude range of 12--17;][]{Bohlin96}.
  The Calspec standards are observed by {\it HST}, an extremely stable
  system where calibration systematic uncertainties are small and well
  understood.  For each star, there is a stellar atmosphere model,
  which can be used to determine the agreement between the
  observations and theoretical expectations.  By comparing different
  Calspec stars, it has been shown that the sample is internally
  consistent (Scolnic et~al., in preparation).  The Calspec standards
  define the absolute flux scale of the PS1 system, and the Foundation
  sample will be the first low-$z$ sample to be tied directly to the
  Calspec standards, completely removing the dependence on Landolt
  standards, Vega, and other calibrators, and thus removing a step in
  our calibration procedure.  Furthermore, because PS1 has been
  cross-calibrated to other high-$z$ surveys \citep{Scolnic15}, any
  future improvement in the calibration of these surveys can directly
  benefit the calibration of the PS1 system.

  Continued PS1 observations will continue to measure any potential
  temporal evolution of the system.  Additionally, several new {\it
    HST} standard stars were specifically chosen to be in the MDFs
  \citep{Narayan16:wd}, meaning that for a subset of standard stars,
  there will be hundreds of observations.

\item {\bf Low- and high-$z$ samples on the same photometric system:}
  Having a large set of both low- and high-$z$ SN~Ia observations from
  a single telescopic system is highly advantageous.  A cosmological
  analysis of such a data set will remove all cross-system calibration
  systematic errors.  As different surveys observe SNe in different
  redshift regimes, calibration systematics between surveys
  potentially introduce large cosmological biases.  The total
  calibration systematic uncertainty for $w$ is currently $4.5\%$.
  This single systematic uncertainty is almost as large as the entire
  statistical error for the recent PS1 analysis
  \citep[5.0\%;][]{Rest14, Scolnic14:ps1}.

  Calibration has been significantly improved over the last 10 years
  for the high-$z$ samples, and their cross-calibration has greatly
  improved their overall calibration \citep{Betoule14, Scolnic17}.
  However, the low-$z$ calibration is only loosely tied to the
  high-$z$ calibration, and a path for improving the current sample is
  not obvious as the calibration systems used for many of the low-$z$
  surveys were not measured using the current state-of-the-art
  techniques \citep{Scolnic15}.  Since some systems have been
  decommissioned, the path to further improvement in those cases is
  not clear.

\item {\bf Established data reduction pipeline:}
  To produce the light curves for the high-$z$ PS1 SN~Ia sample,
  \citet{Rest14} adapted the well-tested {\tt photpipe} data reduction
  pipeline \citep{Rest05:photpipe} to work with PS1 data.  After basic
  data processing by the PS1 Image Processing Pipeline
  \citep[IPP;][]{Magnier06, Magnier13, Waters17}, {\tt photpipe}
  further reduces the data, performs photometry, and generates
  publication-quality SN light curves.  We have made some additional
  minor modifications to this pipeline to process Foundation data.  We
  further describe the data reduction pipeline in Section~\ref{s:obs}.

  Although we are still working to improve this pipeline further
  (e.g., implementing a scene-modeling photometry package;
  \citealt{Holtzman08}), its existence means that any SN observations
  with PS1 can immediately be converted into publication-quality
  photometry {\it with the first SN image.}  Unlike other low-$z$
  surveys, using PS1 avoids the need to expend person power on writing
  or adapting a data reduction pipeline and decreases the time between
  data acquisition and publication of the Foundation sample.

\end{enumerate}

In summary, the PS1 system fulfills all telescopic system requirements
(Section~\ref{ss:tel_req}) and is ideal for generating a new low-$z$
SN~Ia sample for cosmology.

\subsection{Foundation Supernova Survey Strategy}

Having determined that the current low-$z$ SN~Ia sample is impeding
progress in understanding dark energy and that the PS1 system is ideal
for improving the sample, we now describe the Foundation Supernova
Survey strategy.

The Foundation Supernova Survey aims to observe a large sample of
low-$z$ SNe~Ia with PS1.  By observing with PS1, the Foundation sample
will be on one of the best-calibrated photometric systems available,
providing a calibration accurate to a few mmag.  Furthermore, at the
end of the survey, the Foundation Supernova sample will be both large
enough and sufficiently calibrated that adding existing low-$z$
samples to a cosmological analysis will typically provide little
improvement in the resulting measurements.

The Foundation Supernova Survey is a follow-up survey.  The
Pan-STARRS1 Science Consortium finished the 3$\pi$ sky survey in 2014,
and since then the PS1 telescope has been running a wide-area survey
(mostly around the ecliptic) focused on a near-earth object (NEO)
search.  Around 90\% of the time is dedicated to this survey which
takes 4 images each night using \wps\ in dark time and combinations of
\ips and \zps in bright moon time.  These data are used to discover
SNe by the Pan-STARRS Survey for Transients \citep[PSST;][]{Huber15,
  Polshaw15, Nicholl16}, with more than 3000 SN candidates reported to
the IAU Transient Name Server in 2016.  The cadence and filter
deployment of the NEO survey data is not adequate to produce SN light
curves sufficient for measuring distances.  Instead the Foundation
Supernova Survey triggers PS1 to take follow-up observations on known,
low-$z$ SNe, but does use PSST images to supplement our targeted
observations.

\subsubsection{Target Selection}

To avoid biases related with targeted SN surveys, we primarily draw
our sample from untargeted surveys.  At the time of publication, the
majority of (announced) bright ($m < 17$~mag) and faint ($17 < m <
21$~mag) nearby SNe are discovered by the All-Sky Automated Survey for
Supernovae (ASAS-SN) and PS1 through the Pan-STARRS Survey for
Transients (PSST), respectively\footnote{See
  https://wis-tns.weizmann.ac.il/stats-maps for discovery
  statistics.}.  These two surveys are the initial (or secondary,
independent) discoverers of \fracasassnone\% (\fracasassntwo\%) and
\fracpsstone\% (\fracpssttwo\%) of our full sample, respectively.

However, we do not exclusively observe SNe discovered by these
surveys.  Other surveys are similar in cadence and depth and, for
purposes of simulating efficiencies, can be treated as extensions of
these surveys.  Moreover, other surveys, including targeted surveys,
can discover SNe that ASAS-SN and/or PSST would have discovered if
that survey did not exist.  Removing these objects from our selection
could also introduce biases.

To be selected for full follow-up observations, a SN must (1) be in
the Hubble flow ($z > 0.015$) but close enough such that we obtain
high signal-to-noise ratio (S/N) photometric observations with minimal
Malmquist bias ($z \lesssim 0.08$) or close enough where measuring a
Cepheid or Tip of the Red Giant Branch (TRGB) distance is feasible ($D
\lesssim 40$~Mpc), (2) have relatively low Milky Way reddening
($E(B-V)_{\rm MW} < 0.2$~mag), (3) be at $\delta > 30^{\circ}$ to fall
into the 3$\pi$ footprint, (4) be observable by PS1 for at least 45
days, and (5) and be spectroscopically confirmed as a SN~Ia where our
first observation can be scheduled before maximum brightness.

As we require a spectroscopic classification for a full set of
follow-up observations, we are often reliant on public
classifications.  Although we have spectroscopic follow-up time
(primarily with SOAR, SALT, and the KPNO 4-m telescope), other
classifications are often used for selection.  When possible, we
obtain a spectrum of every Foundation SN to confirm its
classification, measure spectral properties, and estimate the phase of
our first photometric observation.

To avoid significant biases related to spectroscopic classification,
we obtain ``snapshot'' observations of potential Foundation SNe before
classification.  These observations consist of a single epoch of
\grips\ data.  SNe for which there is a (sufficiently deep)
non-detection within 15~days of discovery, fulfill other requirements,
and are associated with a $z \lesssim 0.08$ galaxy (including
photometric redshift estimates), and where we think that a public
classification, including from our spectroscopic resources, will
plausibly occur within one week of discovery are observed.  The last
requirement results in observing any reasonable candidate discovered
up to a week before one of our classical spectroscopic telescope
nights (with considerations for declination when our resources are
Southern), any $m < 17$~mag SN discovered after full and before new
moon, and any $m < 15$~mag SN.

We do not perform any additional observations until a classification
spectrum is obtained.  If no classification spectrum is obtained within
a week, we discontinue all observations.  If the spectrum indicates
that the object is not a SN~Ia, it is not within our redshift range,
or our snapshot observation was not before maximum brightness, we
discontinue all observations.

At the current rate, roughly 300 appropriate SNe~Ia are discovered per
year.  The main hurdle is spectroscopic classification.  On a recent
SOAR night, we were able to observe 17 SNe and classify 10 SNe, of
which 6 are now included in the Foundation sample \citep{Pan15:soar}.
Even if only one-third of all SNe observed are ultimately included in
the Foundation sample, we should be able to obtain spectra of \about
150 Foundation SNe per year with two 4-m nights per month.  Such
follow-up time has been awarded for several semesters (PI Foley: NOAO
Programmes 2015A-0253, 2015B-0313, 2017B-0058, 2017B-0169; Lick
Programme 2017A\_S011; Keck Programme 2017A\_U079; PI Jha: SALT
Programmes 2015-1-MLT-002, 2016-1-MLT-007, 2017-1-MLT-002).

We will also make an effort to carefully observe all potential
Cepheid/TRGB SNe~Ia at higher cadence (and shorter exposure times to
avoid saturation and maintain a constant open-shutter time).  Although
this sample will initially be small, we should be able to observe
\about 2 Cepheid/TRGB-SN calibrators per year.  This rate is
relatively small, but current $H_{0}$ analyses use only 19 Cepheid
SNe~Ia \citep{Riess16}, and we will be able to contribute a large
increase in this sample in a few years of observations.  Having the
Cepheid SNe and the Hubble-flow SNe on the same system will remove a
large systematic uncertainty for the measurement of $H_{0}$.  We may
continue to observe Cepheid/TRGB SNe~Ia after the conclusion of the
Foundation Supernova Survey, and increasing statistics now is
important because of the low rate.

\subsubsection{Exposure Time and Cadence}

The primary driver for the exposure time and cadence of the Foundation
Supernova Survey is the statistical distance uncertainty, which we
would like to be smaller than the intrinsic scatter in the sample,
$\sigma_{\rm int}$.  While different studies have found a range of
$\sigma_{\rm int}$ \citep[e.g.,][]{Hicken09:de, Betoule14, Scolnic15},
most have $0.10 < \sigma_{\rm int} < 0.17$~mag.  We therefore aim to
have a statistical distance uncertainty of $<$0.10~mag.

The first consideration for the distance uncertainty is an adequate
cadence to properly measure the peak brightness and light-curve shape
for each SN.  At a minimum, this requires three epochs (before peak,
near peak, and at $t \gtrsim 15$~days).  We performed simulations that
suggest that such a cadence, if performed with sufficiently high S/N
observations is sufficient for measuring the peak brightness and
light-curve shape of a SN~Ia.  However, this is an optimistic
scenario, and cautious observers would generally prefer more
observations.

If we adequately measure the peak of the light curve and the
light-curve shape, the dominant term in the distance uncertainty is
the colour uncertainty.  The intrinsic colour variation for a given
light-curve shape is of order 0.03~mag \citep{Scolnic16, Mandel16},
and therefore, we would like to reach this level of precision.
Additionally, the effect of the colour uncertainty on the distance is
roughly 3 times that of the uncertainty on peak brightness, and to
reach a distance uncertainty of 0.10~mag, we must measure the SN
colour to $\lesssim$0.03~mag.

The colour precision is roughly determined by the overall S/N of all
light-curve points (assuming roughly similar S/N in each band).
Therefore, there is a degeneracy between the exposure time of an
individual epoch and the number of epochs for a light curve.  Here we
examine the optimal values given some constraints.

Our first constraint is that we require $\ge$3 epochs to constrain the
peak brightness and light-curve shape.  Second, the PS1 camera, GPC1,
has a read-out time of 7~s.  To avoid excessive overhead and to make
the images sky-noise dominated (rather than read-noise dominated), we
have limited our exposure times to be a minimum of 15~s.  In a 15-s
exposure, SNe with \grizps\ magnitudes of 18 (20) typically have ${\rm
  S/N} = 44$, 54, 52, and 40 (10, 12, 11, and 8), respectively.  If
observations are set to the minimum exposure time, the light curves
will have sufficient S/N to measure the colour to our desired value in
5~epochs.

However, there are other uncertainties related to the number of
epochs, including the zeropoint uncertainty of local stars, which we
find is typically $<$5~mmag.  This combined with potential
catastrophic events (such as cosmic ray hits or satellite trails) make
a larger number of epochs desirable.  As a result, we aim to obtain
the minimum number of epochs at the shortest exposure time where the
combination results in a colour uncertainty below our goal.  Having
more epochs is also likely more interesting for understanding the
physics of SN~Ia progenitors and explosions and will make our light
curves more useful for light-curve training.  Our simulations have
shown that roughly 7 epochs provides the optimal light curves.  The 7
epochs could be scheduled such that there are different sized gaps
between epochs at different phases of the light curve.  For instance,
one might want to observe nightly near peak and every 10~days later.
However, such a cadence is difficult to obtain in practice.  A shorter
cadence is difficult to consistently achieve given typical stretches
of bad weather, where the telescope can have gaps of $>$3 days.  It
also places additional stress on the telescope scheduling and requires
more effort.  We chose an average cadence around peak of 5~days for
simplicity and to roughly match the cadences of the SDSS and SNLS
surveys.  We chose our final observation to be at $\gtrsim$35~days
after peak to measure the late-time colour (which has been shown to be
a good indicator of dust reddening; \citealt{Lira96}).  The nominal
sequence is then $-5$, 0, 5, 10, 18, 26, and 35~days relative to peak.

For this observation sequence, we have simulated the survey based on
the weather history, sky noise, and depth of the PS1 SN survey
\citep{Scolnic14:ps1} at various exposure times.  We then fit the
simulated light curves and recovered light-curve parameters.  We find
that at our lowest exposure time of 15~s, colour errors are typically
0.03--0.04~mag and distance errors are typically 0.10--0.12~mag, in
line with our goals.

\begin{figure}
\begin{center}
\includegraphics[angle=0,width=3.2in]{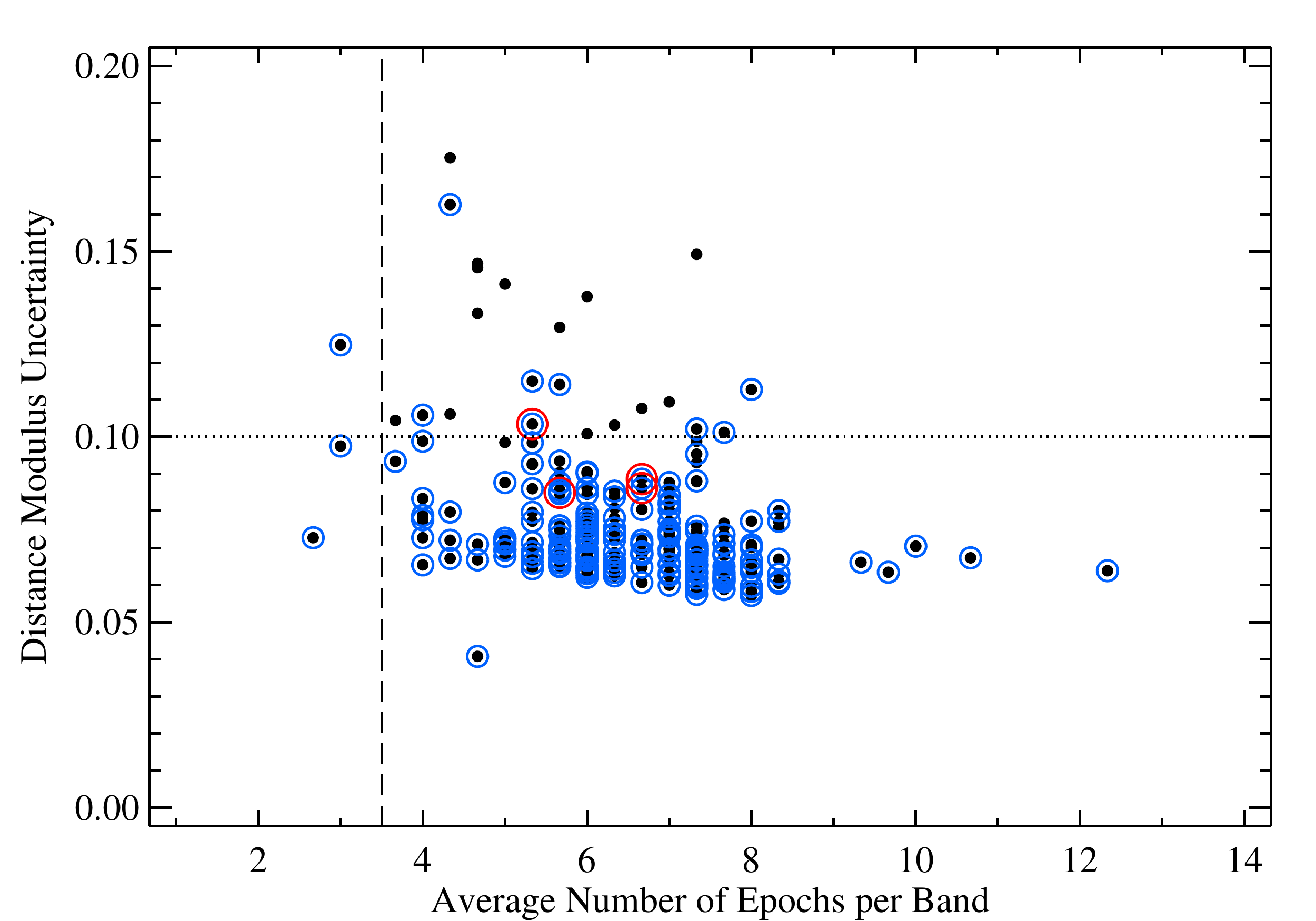}
\caption{Distance modulus uncertainty as a function of average number
  of epochs observed per band for the Foundation sample (black dots).
  Blue circles indicate SNe that pass all criteria for inclusion in
  our cosmology sample (Section~\ref{ss:cuts}) {\it except} for the
  requirement that there be an average of $>$3.6 epochs per band (in
  \grips) and that the first epoch be earlier than 7~days after peak
  brightness.  The red circles indicate SNe that pass all criteria
  {\it except} that the first epoch be earlier than a week past peak
  brightness.  The dotted horizontal line represents our goal of a
  0.10~mag distance modulus uncertainty.  The vertical dashed line
  represents the average number of epochs necessary for inclusion in
  the cosmology sample.  Objects with more epochs have typical
  distance uncertainties of $\le$0.10~mag.}\label{f:epoch}
\end{center}
\end{figure}

\begin{figure}
\begin{center}
\includegraphics[angle=0,width=3.2in]{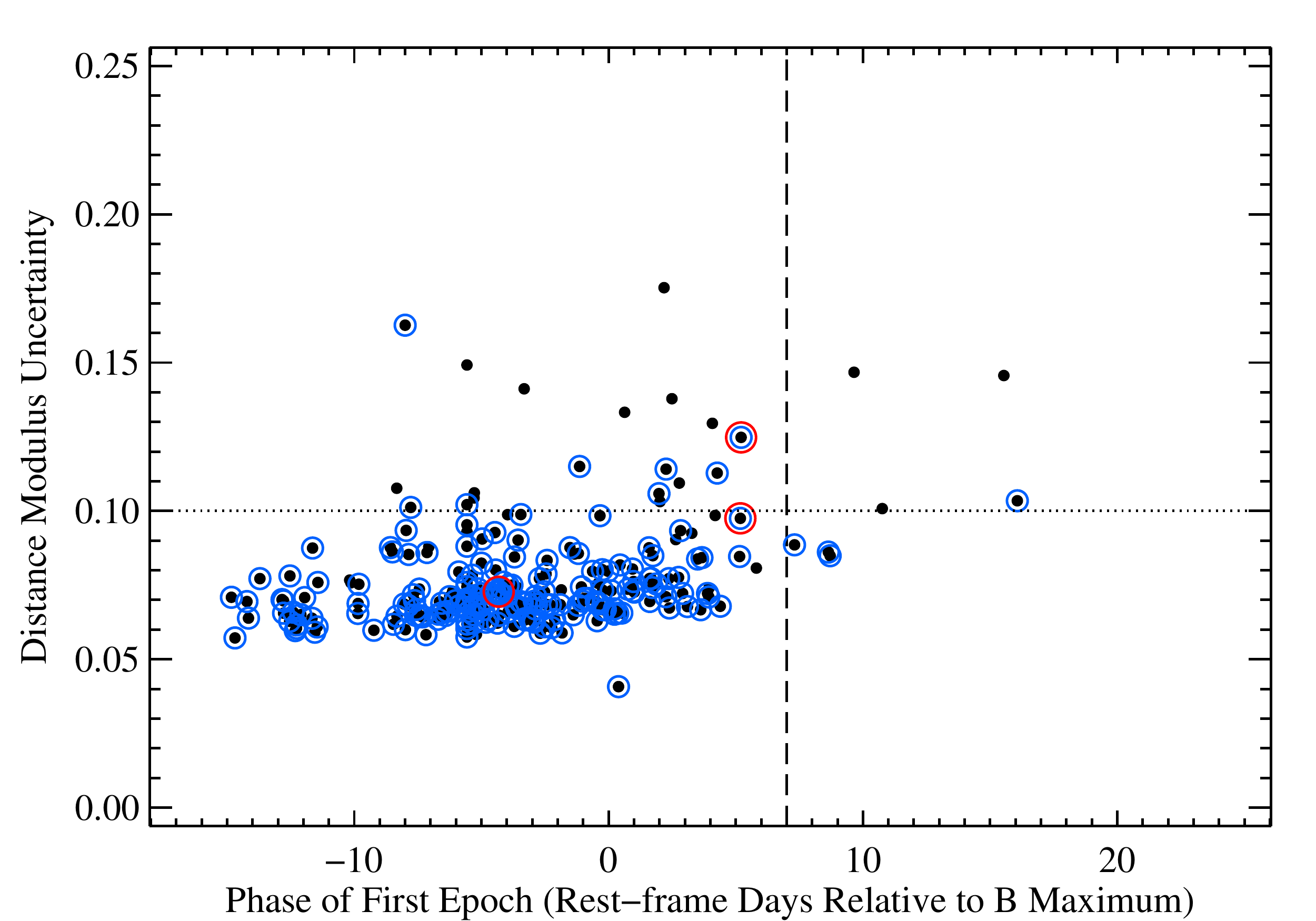}
\caption{Same as Figure~\ref{f:epoch}, except as a function of the
  phase of the first epoch for the Foundation sample.  Here, the red
  circles indicate SNe that pass all criteria {\it except} they have
  too few epochs for inclusion.  The vertical dashed line represents
  the first-epoch phase necessary for inclusion in the cosmology
  sample.  Objects with a first epoch at a phase of $<$+7~days have
  typical distance uncertainties of $\le$0.10~mag.}\label{f:start}
\end{center}
\end{figure}

We have examined this strategy with our survey data (see
Section~\ref{s:obs}).  For each SN observed, we can compare the final
distance error to the number of epochs observed, which we display in
Figure~\ref{f:epoch}.  We note that these are the number of observed
epochs, not the number of epochs used when fitting the light curves,
where particularly early or late data are not included.  We find that
the typical distance uncertainty is $<$0.10~mag for SNe with an
average of $\ge$3.6 epochs per band.  We also find that if a SN is
observed prior to 7 days after peak brightness generally produces
small distance uncertainties (Figure~\ref{f:start}).  For the
Foundation Supernova sample, we find that both criteria are necessary
to have confidence in our distance moduli.

\subsubsection{Data Quality}

We limit our observations to have an airmass below 2, extinction due
to clouds of $<$0.5~mag, and seeing with ${\rm FWHM} < 1.8$\arcsec.
We place each SN at a random place on the focal plane, restricted to
areas known to produce high-quality images (e.g., $>$0.4$^{\circ}$
from the center of the focal plane; \citealt{Rest14}) and away from
known detector defects and chip gaps.  For each SN, the location does
not vary for $>$0.05$^{\circ}$ for the entire light curve.  As a
result, we reduce potential systematic biases related to a particular
detector or position in the focal plane, which is a 3-mmag bias across
the entire focal plane \citep{Scolnic14:ps1}.

The morning after an observation, we check the SN location to make
sure that the nearby area has not been masked, that there are no
artefacts near the SN, and that the SN is sufficiently far from chip
gaps.  If there are any issues with the observation, we immediately
request a new observation.  We have requested repeat observations
because of problems in an image for only \repeatfrac\% of our observations.

For the nights where we requested observations, \weatherfrac\% were
weathered out.  In these cases, we immediately request a new
observation.  Based on our current data set, we have found that our
median cadence near peak is \medcad~days, close to our desired value
of 5~days.  The reason for this difference is primarily caused by
periods of extended bad weather.  Our median seeing is
\medseeg\arcsec, \medseer\arcsec, \medseei\arcsec, and
\medseez\arcsec\ in \grizps, respectively.

Observations typically saturate at 12.5~mag for 15-s exposures.  This
is much brighter than any of our SN observations.  Furthermore, we
expect any non-linearities due to the brighter-fatter effect
\citep{Antilogus14} to be $<$2~mmag at $m > 13$~mag.  The typical
number of stars per CCD that overlap with calibrated stars from the
3$\pi$ survey is $>$500 stars.  Typical uncertainties in our nightly
photometric zeropoints are 3~mmag.

\subsubsection{Sample Size}

For any cosmological analysis using the Foundation sample, the
statistical uncertainties will depend, in part, on the number of SNe
in the Foundation sample.  Since cosmological measurements depend on
comparing the {\it relative} distances to SNe, a large sample of
low-$z$ SNe is critical for precise measurement of cosmological
parameters.  There are currently \about 1000 published high-quality
high-$z$ SNe~Ia \citep{Scolnic17}, about 5 times the size of the
current low-$z$ sample.  As such, a single low-$z$ SN~Ia currently has
more statistical weight than a single high-$z$ SN~Ia.  Moreover,
observations at longer rest-frame wavelengths, where the SNe suffer
less dust extinction and have lower intrinsic scatter
\citep[e.g.,][]{Krisciunas04:hubble, Wood-Vasey07, Folatelli10,
  Mandel11, Barone-Nugent12, Stritzinger11, Friedman15}, are easier to
obtain at low $z$.  Therefore, observing low-$z$ SNe is a relatively
economical approach to improving the statistical uncertainties on
cosmological parameters.

However, gains will steadily decrease as the statistical uncertainty
approaches the systematic uncertainty.  If we can produce a low-$z$
SN~Ia sample that is the same size as the current sample, but with
smaller systematic uncertainties, that is a clear improvement.  The
Foundation sample will have calibration uncertainties similar to that
of the PS1 high-$z$ sample, which are one-third that of the current
low-$z$ sample \citep{Scolnic15}.  As the current low-$z$ sample
contains \about 200 SNe~Ia, a first goal for the Foundation sample is
to match that number.

To demonstrate the power of the reduced systematic uncertainties
associated with the Foundation Supernova Survey, we performed multiple
simulations using the SNANA simulation package
\citep{Kessler09:SNANA}.  The simulations match all key
characteristics of the surveys including cadence, S/N, and
detection/spectroscopic selection functions.  As has been done in past
cosmology analyses \citep{Betoule14, Scolnic14:ps1, Scolnic17}, the
selection functions have been empirically determined such that each
simulation produces the observed redshift distribution as well as the
observed light-curve shape/colour distributions as a function of
redshift.  In the future, we will directly include detection
efficiencies of the ASAS-SN and PSST surveys to better constrain the
overall Foundation Supernova Survey selection function.  Our accurate
simulations will allow us to correct for any distance biases with
redshift, colour and stretch \citep{Scolnic16}.

As a baseline, we simulated the current SN~Ia sample, including the
current low-$z$ sample \citep[but set to have exactly 200 SNe~Ia for
simplicity]{Rest14}, the Joint Light-curve Analysis (JLA) high-$z$
sample \citep{Betoule14}, and the PS1 high-$z$ sample \citep{Rest14}.
SNANA generates light curves as would be observed by each survey that
contributes to the final sample.

Once the Foundation Supernova Survey data are incorporated in SN
spectral models, the reduced calibration uncertainties will improve
the SN spectral model uncertainties by \about 25\% (Scolnic et al., in
prep.), which is roughly the fraction of low-$z$ SNe in the current
training set.  For all simulations, we include the same level of
systematic uncertainties from the absolute calibration of the {\it
  HST} Calspec standards and MW extinction.  No other systematic
uncertainties are included; for instance those related to selection
biases are excluded (although we expect to improve such uncertainties
with the Foundation Supernova Survey in the future).

Cosmological constraints are determined from use of the Markov Chain
Monte Carlo (MCMC) technique implemented with the CosmoMC code
\citep{Lewis00}.  The cosmological parameters are all given
non-informative priors.  For all measurements in this paper, we assume
a flat universe and marginalize over $H_{0}$ and $\Omega_{M}$.

\begin{figure}
\begin{center}
\includegraphics[angle=0,width=3.2in]{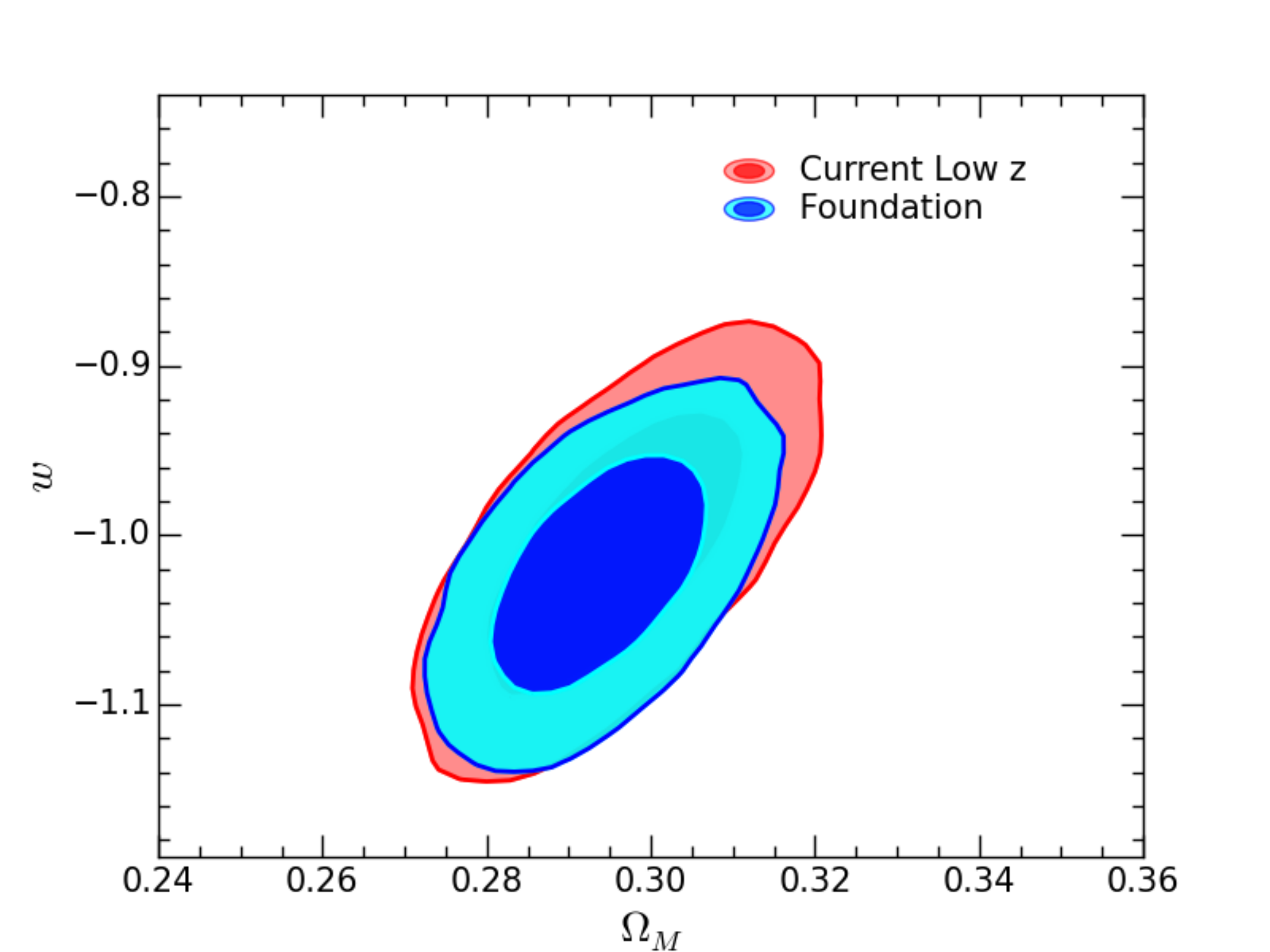}
\caption{1- and 2-$\sigma$ confidence contours in the
  $w$--$\Omega_{m}$ plane for simulated SN data.  The pink/red and
  teal/blue contours represent the projected constraints from current
  CMB \citep{Planck16}, the simulated JLA high-$z$ SN compilation
  \citep{Betoule14}, and either a simulation that roughly matches the
  current low-$z$ SN~Ia sample (but with exactly 200 SNe~Ia) or a
  simulated Foundation Supernova sample (with exactly 200 SNe~Ia),
  respectively.  Including systematic uncertainties, the marginalized
  uncertainty for $w$ improves from 0.053 to 0.046, a 35\% improvement
  in precision, when replacing the current low-$z$ sample with an
  equal-sized Foundation sample.}\label{f:current200}
\end{center}
\end{figure}

When we combine the simulated current low-$z$ SN and JLA samples with
CMB \citep{Planck16} and BAO \citep{Anderson14} data, and assume a
constant dark energy equation of state, we find the uncertainty for
$w$ to be $dw = 0.053$, comparable to that found by \citet{Betoule14},
$dw = 0.055$\footnote{While the simulated results produce slightly
  better results (by 8\% in precision) than the real data, this is
  likely because of the updated CMB constraints used for the
  simulations.}.  We display the $w$--$\Omega_{m}$ confidence contours
derived from these simulated data in Figure~\ref{f:current200}.

We also simulated a sample of 200 Foundation SNe.  The combination of
the simulated Foundation sample, JLA high-$z$ sample, CMB data, and
assuming a constant equation of state, results in a uncertainty for
$w$ of $dw = 0.046$ ($w$--$\Omega_{m}$ confidence contours are shown
in Figure~\ref{f:current200}).  That is, simply replacing the current
low-$z$ SN~Ia sample with the same number of SNe from the Foundation
sample is predicted to result in a 35\% improvement in precision for
our our measurement of $w$.

\begin{figure}
\begin{center}
\includegraphics[angle=0,width=3.2in]{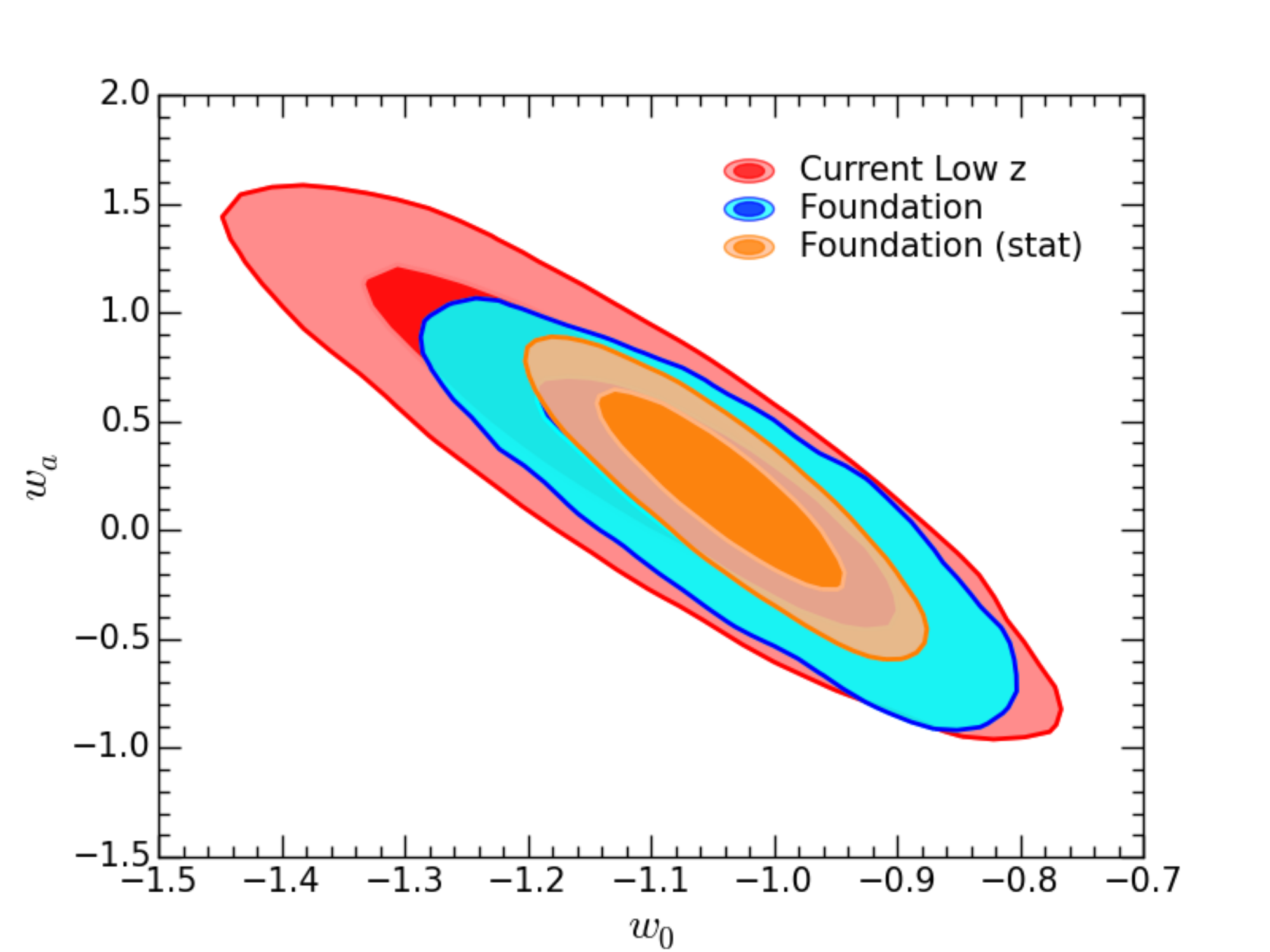}
\caption{1- and 2-$\sigma$ confidence contours in the $w_{0}$--$w_{a}$
  plane for simulated SN data.  The pink/red and teal/blue contours
  represent the projected constraints from current CMB
  \citep{Planck16}, a simulated PS1/DES photometric sample of 3500
  high-$z$ SNe~Ia, and either the current low-$z$ SN~Ia sample or 200
  Foundation SNe~Ia, respectively.  The tangerine/orange contours
  represent what we would expect in the absence of any systematic
  uncertainties.  The dark energy FoM improves from 30 to 52, a 72\%
  improvement, when replacing the current low-$z$ sample with an
  equal-sized Foundation sample.}\label{f:w0wa200}
\end{center}
\end{figure}

In the next few years, significantly larger samples of high-$z$ SNe
(\about 5000 SNe~Ia) from DES \citep{Bernstein12} and PS1
\citep{Rest14} will be available.  With these larger samples, we will
be able to place relatively tight constraints on evolving dark energy.
We simulated 3500 high-$z$ SNe and set the calibration systematic
uncertainty to be the same as PS1.  Combining these data with the
simulated, current low-$z$ and Foundation Supernova Survey samples
(and now allowing for a evolving equation of state), we find dark
energy FoMs of 30 and 52, respectively (Figure~\ref{f:w0wa200}).  That
is, replacing the current low-$z$ sample with an equally sized
Foundation Supernova sample will improve the dark energy FoM by 72\%.

As shown from our simulations, a Foundation sample of \about 200
SNe~Ia is well motivated.  However, larger sample sizes, up to at
least 800 SNe, will also significantly improve cosmological inference
{\it if} potential calibration systematics are appropriately reduced.

In their final report, the {\it WFIRST} Science Definition Team
\citep[SDT;][]{Spergel15} determined that a low-$z$ sample of 800
SNe~Ia is {\it required} to reach their science goals.  While this is
further confirmation of the importance of a large high-fidelity
low-$z$ SN~Ia sample, it also presents an independently derived number
for a final sample size.  While 200 or 400 Foundation SNe~Ia would
potentially be a significant fraction of the final {\it WFIRST}
low-$z$ required sample, it is possible for the Foundation sample to
be the entire {\it WFIRST} low-$z$ sample within 4 years of
operations.

Detailed simulations of the {\it WFIRST} survey are underway
\citep{Hounsell17}, and we will revisit the necessary low-$z$ sample
size in the future.




\section{Observations and Data Reduction}\label{s:obs}

As of May 2017, we have observed a total of \ntot\ SNe with PS1.  Of
these, \nsnap\ were snapshot observations that were not continued.  We
have followed \nsn\ SNe~Ia, whose light curves are presented below.
Most observations are a series of 15-s \grizps\ exposures.  Our
earliest observations were obtained in twilight as a pilot programme
and had 100-s exposures.

We reduce the Foundation PS1 data with the same custom-built pipeline
as for the PS1 MDF survey data.  The basic data processing is
performed by the PS1 IPP \citep{Magnier06, Magnier13, Waters17}.
Down-stream processing is performed with the {\tt photpipe} pipeline
that members of our team developed for the SuperMACHO and ESSENCE
surveys \citep{Rest05:photpipe, Miknaitis07, Rest14, Scolnic14:ps1,
  Narayan16:essence}.

\begin{figure*}
\begin{center}
\includegraphics[angle=0,width=3.4in]{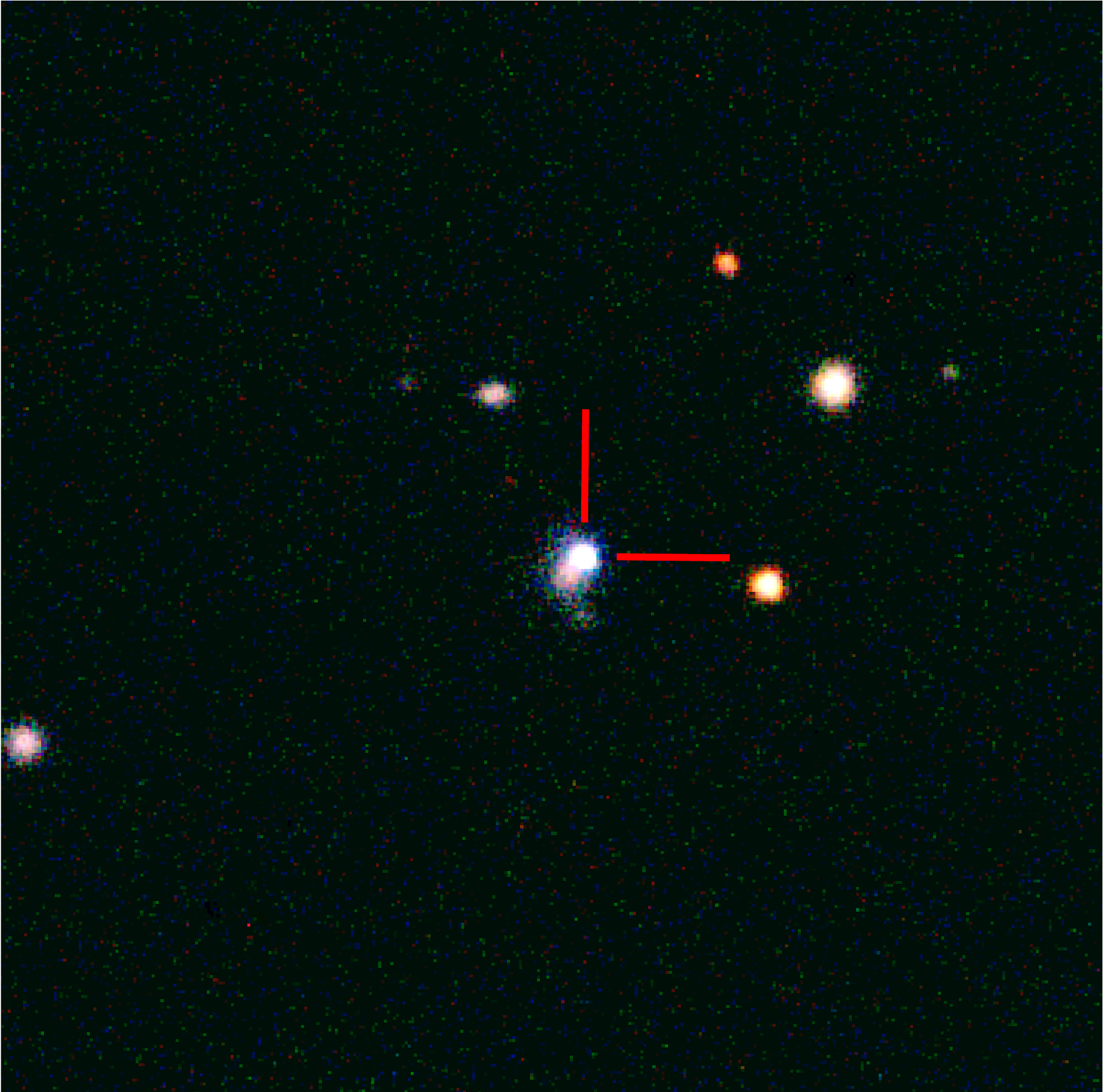}
\includegraphics[angle=0,width=3.4in]{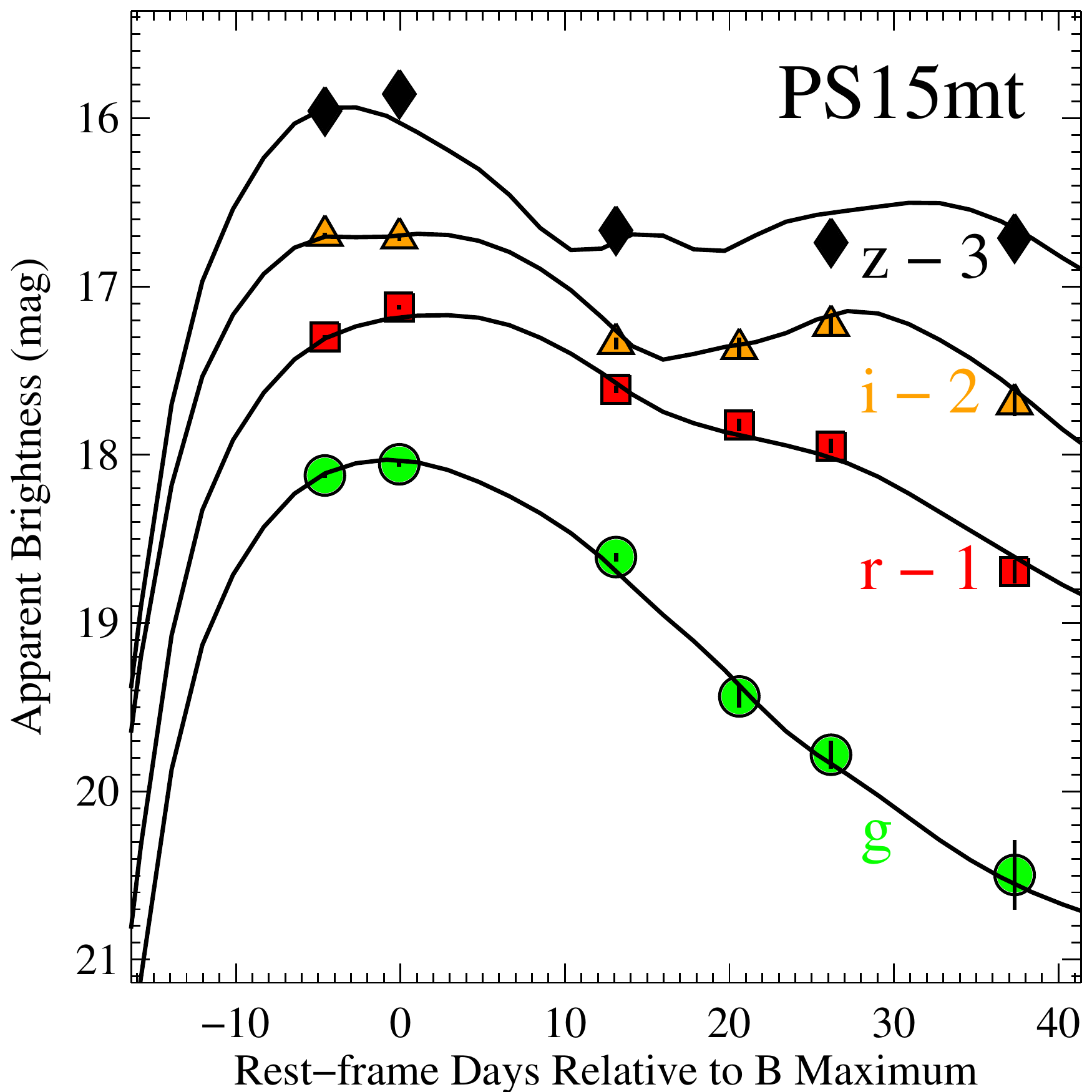}
\caption{({\it Left}) 90\arcsec\ $\times$ 90\arcsec\ PS1 \gri\
  (corresponding to BGR channels) image of PS15mt. The image is
  oriented with North up and East left.  The SN is marked by the red
  tick marks. ({\it Right}) Multi-colour light curves for PS15mt.
  \grizps\ photometry are plotted as green circles, red squares,
  orange triangles, and black diamonds, respectively.  Photometric
  uncertainties are plotted, but are typically smaller than the
  points.  The black lines are SALT2 model fits to the light curves.
  The $z$-band fits are only illustrative; those data are not
  currently used to measure distances.}\label{f:single}
\end{center}
\end{figure*}

The PS1 IPP system performs flat-fielding on each individual image,
using white-light flat-field images of a dome screen, in combination
with an illumination correction obtained by rastering sources across
the FOV.  After determining an initial astrometric solution
\citep{Magnier08, Magnier17}, the flat-fielded images are then warped
onto the tangent plane of the sky, using a flux-conserving algorithm.
We present an example SN image in Figure~\ref{f:single}.

High-quality images of a given SN location obtained during the 3$\pi$
survey are stacked, allowing for the removal of defects such as cosmic
rays and satellite streaks.  Since the 3$\pi$ survey finished in 2014,
these images are free of SN flux for the Foundation sample.  The
stacked images typically go much deeper than the Foundation Supernova
Survey data, with typical limiting magnitudes of 23.0, 22.9, 22.9, and
22.2 in \grizps, respectively \citep{Chambers17}.

Single-epoch Foundation Supernova Survey images are then processed
through a frame-subtraction analysis using {\tt photpipe}.  This
robust and well-tested system first determines the appropriate
spatially varying convolution kernel using {\tt HOTPANTS}
\citep{Becker15}; this kernel is necessary to match match the template
image to the survey image.  After the convolution is performed, the
template image is subtracted from the survey image.  We then detect
significant flux excursions in the difference images using {\tt
  DoPHOT} \citep{Schechter93}.

For each SN, we calculate the weighted average position using
detections from all bands.  Because almost all of the SNe are measured
with high SNR ($>$50) near peak and each have \about 5 exposures, the
typical uncertainty of the SN location is $<$0.1 pixels.  We perform
``forced'' photometry for this position for all epochs for a
particular SN.  This is the same method as used for \citet{Rest14},
with the only difference being that we currently use DoPHOT instead of
DAOPhot.  The median difference between forced and unforced photometry
is \about 1.5~mmag for the brighter detections, and is slightly larger
for low-S/N detections.  The latter difference is the result of an
Eddington bias in the flux of measurements detected without forced
photometry.

Zeropoints are determined from comparing measurements of stars in the
survey image to those in the 3$\pi$ catalog.  The median zeropoint
uncertainty is \about 3~mmag per band.  We present example light
curves for a single SN in Figure~\ref{f:single}.
Eventually, we plan to re-reduce the photometry with a process similar
to that of {\tt DAOPhot} \citep{Stetson87} with forced photometry on
all subtracted images.  This process has been done for the high-$z$
PS1 data \citep{Rest14}, including modifications to minimize
systematic uncertainties in the photometry.

For the Foundation Supernova Survey, we must carefully measure the
system throughput as the filters may have changed since the initial
PS1 calibration was performed.  During the 1.5 years of the initial
PS1 survey, no changes were detected to a limit of 3~mmag
\citep{Rest14}.  This analysis will be extended for the entire
4.5-year PS1 survey.  However, for the Foundation Supernova Survey,
which will observe well past the initial PS1 survey, we plan to
continue to monitor the system throughput.  The majority of PS1
observations are not for the Foundation Supernova Survey, but every
\grizps\ image that PS1 observes provides a direct comparison for
hundreds of stars.  We will use these data, along with specific
calibration frames, to track the system throughput for the duration of
the Foundation Supernova Survey.

\begin{figure*}
\begin{center}
\includegraphics[angle=0,width=6.8in]{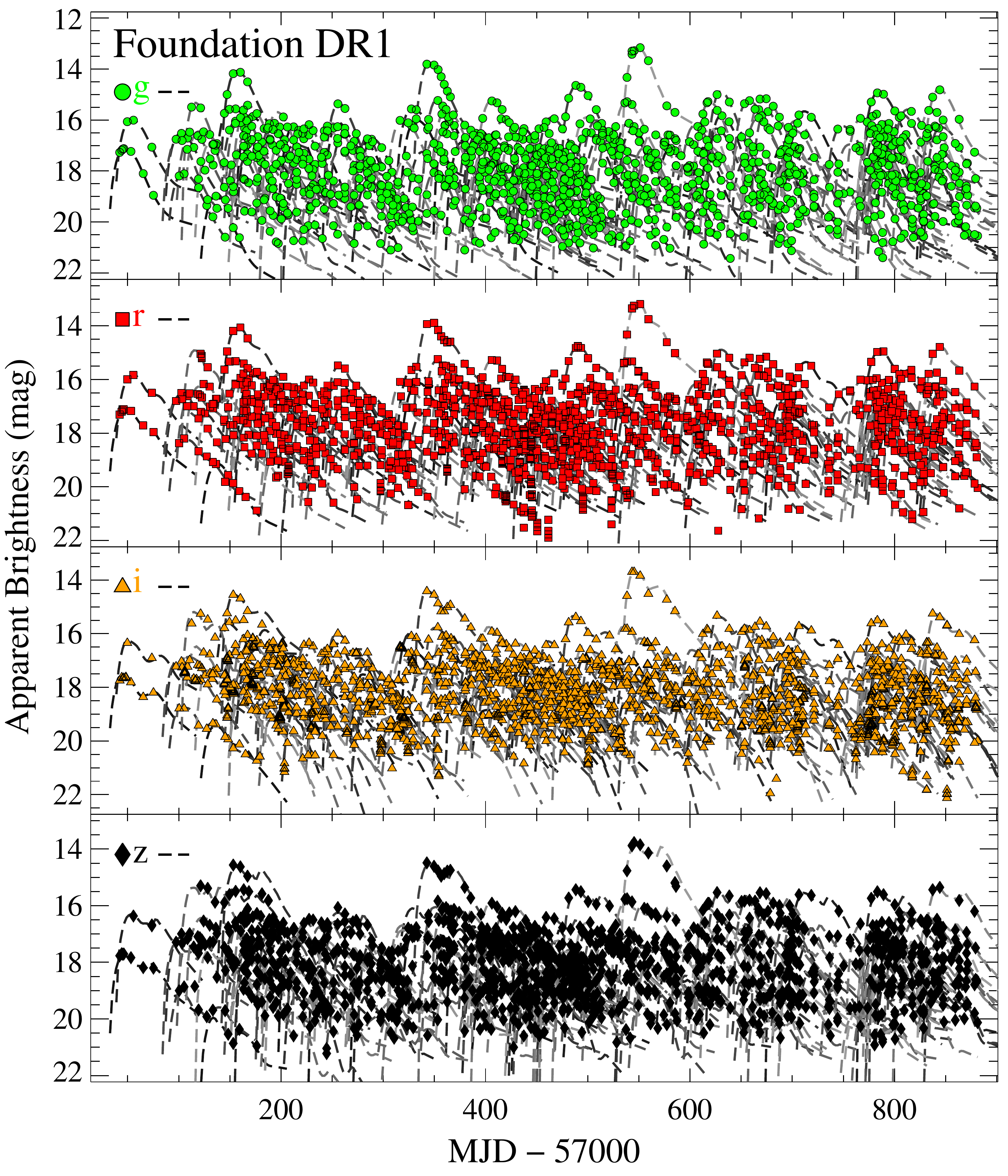}
\caption{Multi-colour light curves for the Foundation DR1 sample.  The
  \grizps\ photometry are plotted as green circles, red squares,
  orange triangles, and black diamonds, respectively.  The grey lines
  are SALT2 model fits to each SN light curve.}\label{f:all}
\end{center}
\end{figure*}

Detailed information about the SNe in our sample is presented in
Table~\ref{t:sample}.  We present photometry of our current sample,
corresponding to the first data release (DR1) of the Foundation
Supernova Survey, in Table~\ref{t:lc} and display their light curves
in Figure~\ref{f:all}.

\section{Additional Data}\label{s:add_data}

If the Foundation sample is to become the benchmark low-$z$ SN~Ia
sample, it must include data beyond the PS1 SN photometry.  In
particular, SN spectroscopy, SN NIR photometry, and host-galaxy data
are all potentially important.

SN spectroscopy is critical for proper SN classification.  While
public classifications are generally reliable, some are incorrect.  We
therefore attempt to obtain our own spectrum of every Foundation SN to
verify its classification.  For some SNe, these data will also provide
the first spectroscopic redshifts of their host galaxies.  For $z <
0.08$, redshift uncertainties can propagate to a relatively large
distance uncertainty.  For instance, at $z = 0.015$, \medz, and 0.08
(our nominal minimum, median, and maximum redshifts for the Foundation
sample), a redshift uncertainty of 0.01 propagates to a distance
modulus uncertainty of 2.40, \medmuerr, and 0.15~mag, respectively.

SN spectral data also have the ability to significantly improve SN
distance estimates \citep[e.g.,][]{Bailey09, Blondin11, Chotard11,
  Foley11:vel, Foley11:vgrad, Foley12:vel, Silverman12:lc, Mandel14}.
Specifically, the intrinsic colour of a SN~Ia is correlated with its
near-maximum light ejecta velocity.  Since there is a broad and skewed
SN~Ia intrinsic colour distribution, an incorrectly assumed intrinsic
colour will result in an incorrect reddening estimate, and thus an
incorrect distance measurement.  Therefore measuring the ejecta
velocity removes a potential bias in cosmological analyses.  While it
is unlikely that every Foundation SN will have a measured near-maximum
ejecta velocity, precisely measuring the distribution is sufficient
for debiasing all SNe in the sample (however those with velocity
measurements will have more precise distances).

SN host-galaxy data can improve distance measurements and reduce a
potential systematic bias.  After making all light-curve shape
corrections, there remains a trend between Hubble residuals and
host-galaxy properties such as stellar mass and metallicity
\citep[e.g.,][]{Kelly10, Lampeitl10:host, Sullivan10, Pan14}.  Since
the Foundation Supernova Survey draws its sample primarily from
untargeted surveys, there is no inherent bias towards massive host
galaxies (unlike previous low-$z$ samples).  Recent cosmological
analyses have removed this trend \citep[e.g.,][]{Betoule14}; however,
the exact size of the effect, its functional form with host-galaxy
parameters such as mass (e.g., a step function, a linear trend, etc),
if there is evolution with redshift, and its physical cause are all
currently unknown.  None the less, it is clear that having these data
are critical for making the best cosmological measurements.

With PS1, we already have sufficiently deep \grizyps\ imaging for
nearly every host galaxy.  From these data alone, we can determine
properties such as stellar mass and SFR to reasonable precision.  To
these data, we will add GALEX UV, 2MASS NIR, and WISE IR measurements
when possible.  For the current Foundation sample, the vast majority
of host galaxies have detections and/or constraining limits in these
data sets.  With these broad-wavelength coverage SEDs, the Foundation
host galaxies will have very precisely measured parameters.

NIR light curves provide a path to smaller distance scatter than
obtained with only optical light curves \citep{Mandel09, Dhawan17}.
This is because of a combination of lower dust extinction (resulting
in a smaller error from extinction corrections) and a theoretically
predicted smaller luminosity scatter in these bands
\citep{Kasen07:wlr}.  While the \yps\ band, with a central wavelength
of 0.96~$\mu$m, technically covers some NIR wavelengths, observations
in \jhk\! bands both extend the lever-arm for extinction measurements
and provide information that is relatively uncorrelated from the
optical \citep{Mandel11}.  We have begun coordinating with multiple
groups to obtain NIR light curves of a subset of Foundation SNe.
Although only a relatively small subset of Foundation SNe will have
NIR light curves, this subsample may be particularly important for
determining the reddening distribution of the full sample as well as
being a high-fidelity training sample for {\it WFIRST}.


\section{Results}\label{s:results}

\subsection{Classifications and Redshifts}

Every SN presented in DR1 has been classified as a SN~Ia.  References
for initial classifications are listed in Table~\ref{t:sample}.  For
those objects where we have access to a spectrum, either from our own
observations, through public surveys such as the Public ESO
Spectroscopic Survey for Transient Objects Survey
\citep[PESSTO;][]{Smartt15}, or through services such as the Transient
Name Server (TNS), we examined each spectrum individually.  Usually,
the initial classifications are consistent with our classifications,
but on occasion there are differences in subclassification and/or
redshift determination.  We present those objects and the differences
below.

Since the Foundation SN sample is drawn primarily from untargeted
surveys, many SNe in our sample do not have a catalogued host-galaxy
redshift.  To determine the redshifts for all objects, we take a
tiered approach.  First, we use redshifts from public surveys with
precise redshifts, which is possible for \about 70\% of all SNe in our
sample.  Second, we use our redshifts measured from our own spectra of
the host-galaxy nucleus.  Third, we use redshifts determined from our
own spectra host-galaxy features at the SN position (usually present
in the SN spectrum).  Fourth, we use redshifts from host-galaxy
features from public SN spectra.  Fifth, we use the SN spectra to
determine a redshift.  Finally, if necessary, we use redshifts
reported by other groups.  We detail this process below.

For all objects where there was an easily identified host galaxy, we
used the NASA/IPAC Extragalactic Database (NED) to determine if there
is a host-galaxy redshift.  For those with multiple redshift
measurements, we chose the one with the smallest uncertainty.  A small
subset of objects (e.g., ASASSN-15nq) have host-galaxy redshifts in
other catalogues (for ASASSN-15nq, the galaxy has an SDSS spectrum),
and we use those sources when necessary.

When no catalogued galaxy redshift is available, we first use our own
host-galaxy spectra, which are generally obtained when observing a SN,
to measure a redshift.  Occasionally, host-galaxy emission or
absorption features are present in the SN spectrum itself, and we are
able to measure a redshift from those.  When possible, we obtain our
own spectrum of the host-galaxy nucleus, often after the SN has faded.
Rarely, other groups will report an otherwise unknown host-galaxy
redshift when classifying a SN, and we use those data when
appropriate.

After all attempts to obtain a host-galaxy redshift are made, there
remain a number of SNe for which we must rely on the SNe themselves
for a redshift (currently 5 SNe, or 2.3\% of the sample).  When we
have access to a spectrum, we use the SuperNova IDentification (SNID)
software \citep{Blondin07} to cross-correlate with SN~Ia templates and
determine the redshift \citep[see e.g.,][]{Foley09:year4, Rest14}.

As a last resort, we use redshift estimates from SN spectra as
presented in classification announcements.  For all direct
measurements, we either use the reported redshift uncertainties or
measure them ourselves.  For reported host-galaxy and SN redshifts, we
assume uncertainties of 0.001 and 0.01, respectively.  We present
newly measured redshifts in Table~\ref{t:redshift}.  In total, the
host galaxies for \ngal\ SNe~Ia in our sample (\ngalper\%) did not
have a catalogued redshift, a surprisingly high fraction.

All SN spectra obtained by our group will be presented in a future
analysis.  For the subset of SNe where either our redshift or
classification differs substantially from the initial classification,
we present those differences in Tables~\ref{t:zdiff} and
\ref{t:cdiff}, respectively.  For 21 (4) SNe, corresponding to 9.3\%
(1.9\%) of our sample, the differences between the initially reported
redshift and our adopted redshift is large enough to produce a
distance modulus bias of $>$0.3~mag ($>$1~mag).  Host-galaxy redshifts
are critical to reduce the scatter of low-$z$ SN~Ia samples; the
Foundation Supernova Survey is currently 97.7\% complete.

\subsection{Light-curve Parameters}

While there are currently several algorithms for estimating distances
from SN light curves (e.g., MLCS, \citealt{Jha07}; SiFTO,
\citealt{Conley08}; BayeSN, \citealt{Mandel09, Mandel11}; SNooPy,
\citealt{Burns11}; BaSALT, \citealt{Scolnic14:col}), currently the
most widely used light-curve fitting algorithm is SALT2
\citep{Guy07}.  This algorithm is based on SALT \citep{Guy05} and
assumes a \citet{Tripp98} parameterization,
\begin{equation}
  \mu_{B} = m_{B} - M + \alpha x_{1} - \beta c, \label{e:tripp}
\end{equation}
where $\mu_{B}$, the distance modulus, is determined for each SN given
observables $m_{B}$, the peak $B$-band brightness, $x_{1}$, a
light-curve shape parameter, and $c$, an observed colour parameter.
The parameters $M$, $\alpha$, and $\beta$ are nuisance parameters that
are globally fit.  One can consider $\alpha$ and $\beta$ to be the
linear slopes between absolute magnitude and light-curve shape and
colour, respectively.  Meanwhile, $M$ is the $B$-band absolute
magnitude of a fiducial SN~Ia with $x_{1} = c = 0$.

We use the most recent version of SALT2 \citep{Guy10} as implemented
in SNANA \citep{Kessler09:SNANA}.  When fitting light curves, we
follow the procedure of \citet{Rest14}, and we refer the reader there
for more details.  After an initial fit, we remove any suspicious
photometry points from each light-curve.  We identify such points as
having $\chi^{2} > 10$ relative to the best-fitting SALT2 model.  This
cut is similar to that done by \citet{Holtzman08} and \citet{Rest14}.
We visually inspected many of the epochs with bad photometry and found
a high correlation with image streaks and other subtraction artefacts,
indicating that these data are generally poor rather than an issue
with the data reduction or model.  In total, 81 out of 6868 data
points (1.2\%), averaging 0.36 data points per SN, were removed.  We
present the parameters which result from light-curve fitting in
Table~\ref{t:param}.

\subsection{Sample Cuts}\label{ss:cuts}

When targeting specific SNe, we only required that the SN be
spectroscopically confirmed as a young SN~Ia with $0.015 < z \lesssim
0.08$.  As a result, we include some SNe~Ia that are spectroscopically
peculiar and have relatively high host-galaxy reddening.  We are
hopeful that these SNe will be cosmologically useful in the future,
but such SNe are not well fit by SALT2, and must be culled from our
current cosmology sample.

Our cosmology sample is set using a combination of criteria defined by
\citet{Rest14} and \citet{Betoule14}.  We list the criteria below, but
refer the reader to previous works for additional details.  The
criteria are:
\begin{enumerate}
\item The SN is not spectroscopically similar to SNe~Iax
  \citep{Foley13:iax, Jha17}, SN~1991bg \citep{Filippenko92:91bg,
    Leibundgut93}, the peculiar SN~2000cx \citep{Li01:00cx}, and the
  high-luminosity SN~2006gz \citep[e.g.,][]{Howell06, Hicken07,
    Yamanaka09, Silverman11:09dc, Taubenberger11, Scalzo12}.
\item The Milky Way reddening towards the SN is $E(B-V)_{\rm MW} <
  0.25$~mag.
\item At least 11 total light-curve points in \grips.
\item The first light-curve point has a phase of $<$+7~days.
\item The uncertainty on $x_{1}$ is $<$1.
\item The uncertainty on the time of peak brightness is $<$1~day.
\item $-0.3 < c < 0.3$.
\item $-3.0 < x_{1} < 3.0$.
\item Chauvenet's criterion applied to the pulls (rather than the
  residuals, which would bias against the lowest-redshift SNe with
  larger peculiar velocity scatter) to exclude systematic outliers.
\end{enumerate}
The first criterion is to remove spectroscopically peculiar SNe which
may be poorly fit by SALT2 or do not follow the width-luminosity
relation \citep{Phillips93}.  The next three criteria are light-curve
quality cuts, which help ensure that the resulting parameters are
reasonable and reduce the number of unreliable light-curve fits.  The
fourth and fifth criteria are the bounds of the SALT2 model.  The
final criterion excludes systematic outliers that are far from a
normal distribution.  The SNe passing all of these criteria compose
the Foundation ``cosmology'' sample.  Table~\ref{t:cuts} displays how
each criterion affects our sample as well as the cumulative effect of
the criteria.

Upon closer inspection of spectra for our sample, we have determined
that PS15zn is likely a SN~Ic rather than a SN~Ia
\citep{Pan15:ps15zn}.  We therefore exclude the SN from our cosmology
sample, but still include its photometry here.

We have decided to observe SNe similar to SNe~1991bg and 2006gz,
knowing that SALT2 cannot properly fit their light curves.  We note
that for the former case, this is not because the SNe are inherently
uncalibratable; for instance, MLCS \citep{Riess96, Jha07} is able to
obtain reasonable distances for SN~1991bg-like objects.  Rather than
potentially bias the sample because of a classification by other
groups or our current ability to properly estimate distances to
particular subclasses of SNe~Ia, we have decided to photometrically
follow this small subset of SNe, and remove them from any current
cosmological analysis.  Based on public classifications and our own
spectra, we designate ASASSN-15ga, MASTER OT J222232.87-185224.3,
PSN~J16283836+3932552, SN~2016ajf, and SN~2016arc as similar to
SN~1991bg and ASASSN-15hy, PSN~J23102264+0735202, SN~2015M,
SN~2016alt, 2016ccj, and 2017mu as similar to SN~2006gz.  These
objects represent 5.3\% of our total sample, but notably, ASASSN-15ga,
ASASSN-15hy, and MASTER OT J222232.87-185224.3 would have been
excluded by other criteria.

Only three objects were cut because of insufficient light-curve
coverage.  One, ASASSN-15bm, was observed during our pre-survey stage,
making it a special case.  Of the remaining criteria, none affects
more than 5\% of the sample.  In total, we lose 15.6\% of our sample
to various cuts, but only 10.8\% of the ``normal'' SNe~Ia.  While we
hope to improve this number as the survey progresses, the Foundation
Survey is already extremely efficient relative to other low-$z$
surveys.  For instance, 54\%, 54\%, and 47\% of the CfA3, CfA4, and
CSP samples \citep{Hicken09:lc, Hicken12, Contreras10} were cut from
the \citet{Rest14} cosmology sample.  Rather, the Foundation sample
has a similar fraction of SNe excluded from high-$z$ samples such as
SDSS \citep[26\%;][]{Sako14, Betoule14} and PS1
\citep[24\%;][]{Rest14}.

\subsection{Sample Demographics}

\begin{figure}
\begin{center}
\includegraphics[angle=0,width=3.2in]{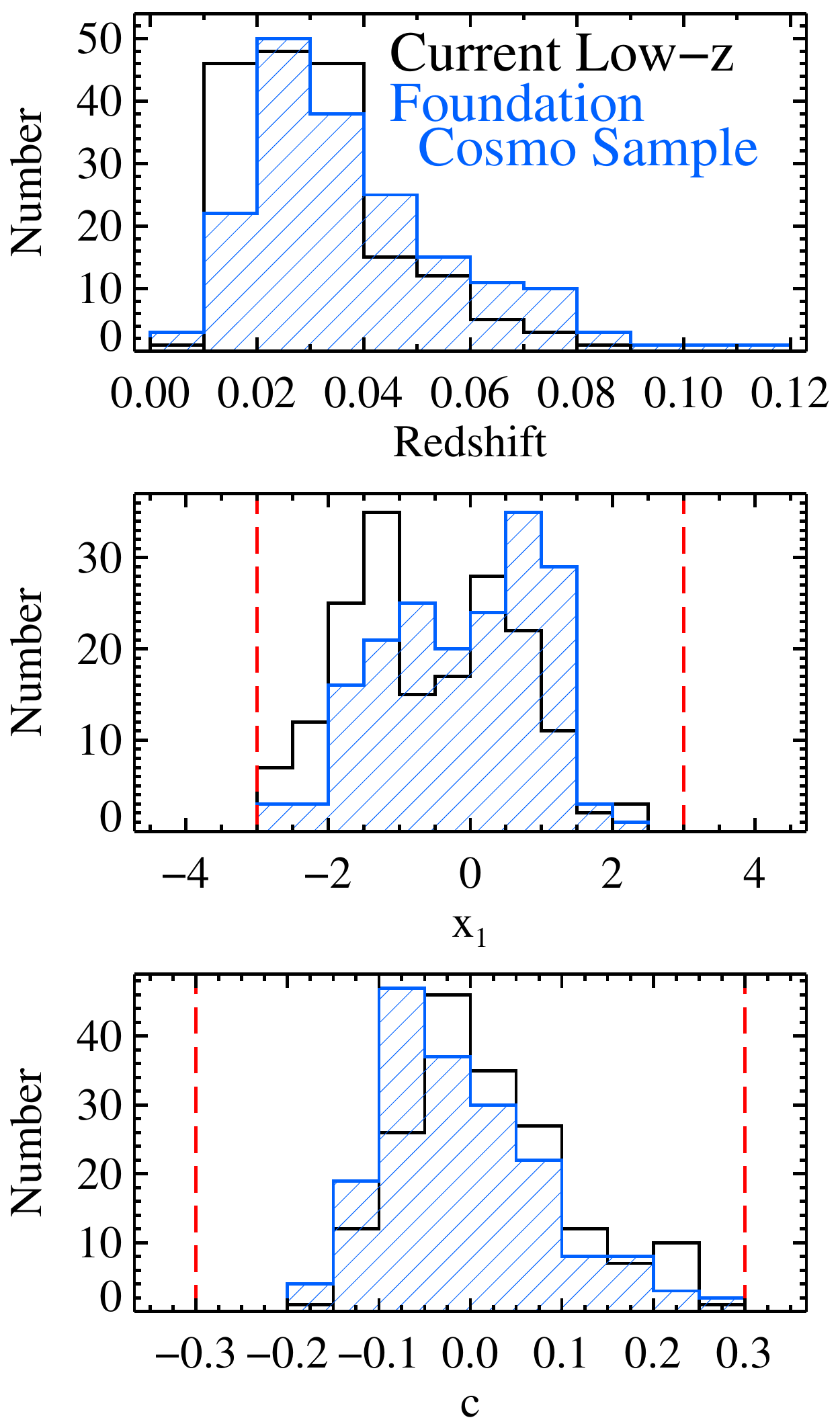}
\caption{Distributions of redshift (top panel), $x_{1}$ (middle), and
  $c$ (bottom) for the Foundation cosmology (blue hashed), and
  existing low-$z$ (black) SN samples.  The vertical dashed lines
  correspond to the parameter limits for inclusion in the cosmology
  sample.  The median values for the Foundation cosmology (existing
  low-$z$) sample are \medz\ (0.029), \medx\ (0.160), and \medc\
  ($-0.035$), respectively.}\label{f:hist}
\end{center}
\end{figure}

Figure~\ref{f:hist} displays the redshift, light-curve shape
($x_{1}$), and observed colour ($c$) distributions of the Foundation
sample compared to the existing low-$z$ sample (as prepared by Scolnic
et~al., in prep.).  For the Foundation sample, we present the results
for the full sample and the cosmology sample, which pass the various
criteria listed in Section~\ref{ss:cuts}.

The low-$z$ SN light curves were fit with the same version of SALT2
and values for the nuisance parameters to give a consistent
comparison.  We note that the low-$z$ sample has already been culled
of SNe that do not pass various quality cuts which are similar to
those employed for the Foundation sample.  The Foundation cosmology
sample has relatively similar demographics to the existing low-$z$
cosmology sample.  The Foundation cosmology and existing low-$z$
samples have similar median redshifts (\medz\ compared to 0.029) and
median colour (\medc\ compared to $-0.035$).  However, the typical
light-curve shapes are different with median values of $x_{1} =
-0.203$ and 0.160 for the Foundation and existing low-$z$ sample,
respectively.  The Foundation $x_{1}$ distribution does not have the
double-peaked distribution seen in the existing low-$z$ sample.

The most striking feature is the relative excess of slow-declining SNe
for the Foundation sample.  While this may be caused by selection
effects (of either Foundation or the other low-$z$ surveys), issues
with fitting the Foundation light curves, or statistical fluctuations,
it does not appear to be a significant issue for measuring
cosmological parameters.  Regardless, we will re-visit this issue with
future data releases.

We note that 18.4\% and 3.0\% of the Foundation SN sample has been
classified as being similar to SN~1991T (or SN~2006gz) and SN~1991bg,
respectively.  These fractions are similar to the magnitude-limited
fractions found by \citet{Li11:rate2} of 17.7\% and 3.3\%,
respectively.

\subsection{Hubble Diagram}

\begin{figure*}
\begin{center}
\includegraphics[angle=0,width=6.8in]{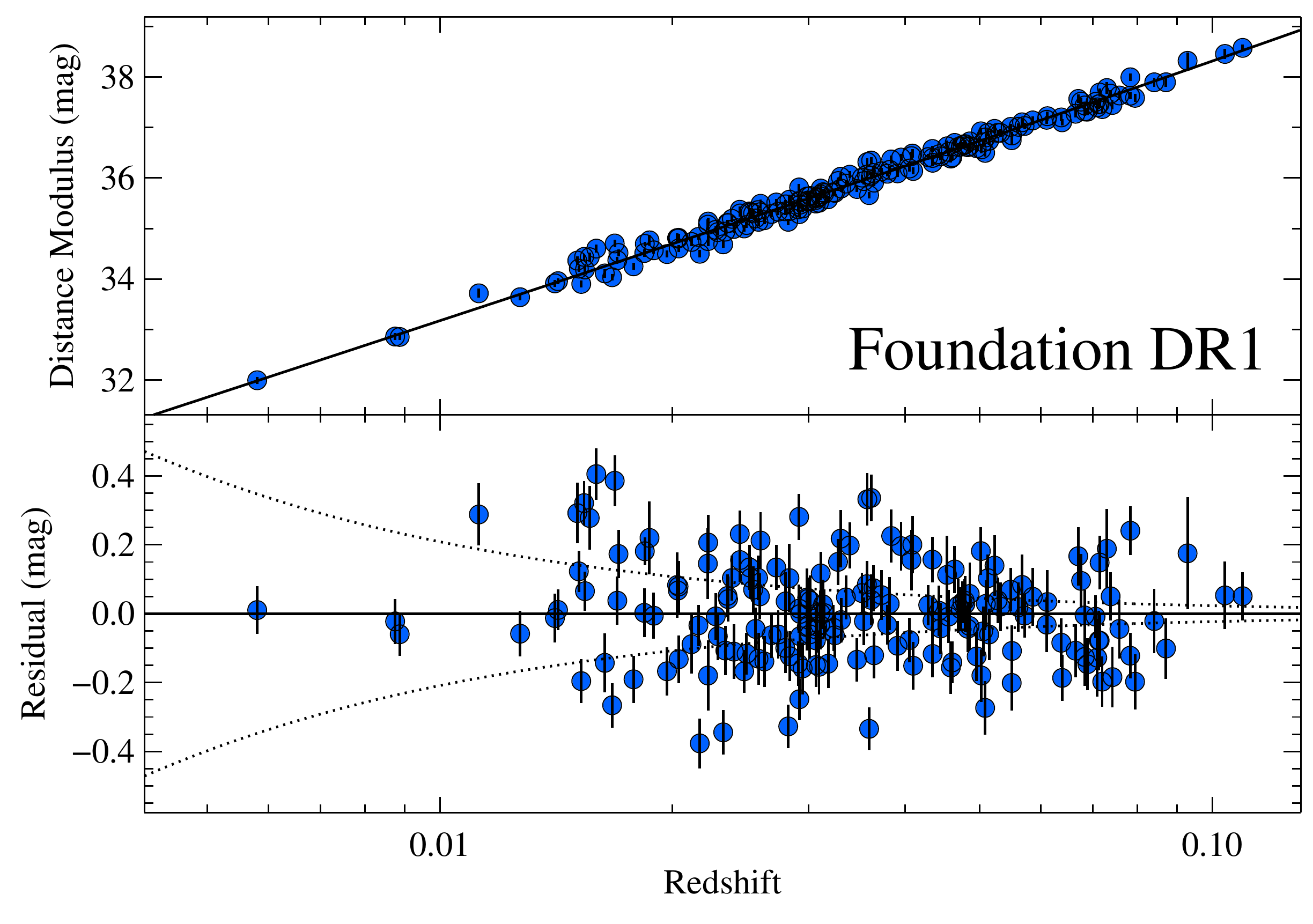}
\caption{Hubble diagram for the Foundation DR1 sample and residuals to
  a fiducial $\Lambda$CDM model (lower panel).  Error bars do not
  include uncertainties related to peculiar velocities (which are
  represented by the dotted curves in the lower panel) and redshift
  uncertainties.}\label{f:hubble}
\end{center}
\end{figure*}

Using the measured distance moduli and redshifts for the Foundation
cosmology sample, we present a Hubble diagram in
Figure~\ref{f:hubble}.  Although the scatter in the Hubble diagram is
small, we caution using the current data in more detailed cosmological
analyses at this time.  We have not yet produced a robust systematic
analysis or determined accurate distance bias corrections (e.g., from
Malmquist bias; these are typically up to \about 0.02~mag;
\citealt{Scolnic17}).  Similarly, we have not measured the necessary
host-galaxy properties to obtain the precise distances required for
such work.

\begin{figure}
\begin{center}
\includegraphics[angle=0,width=3.2in]{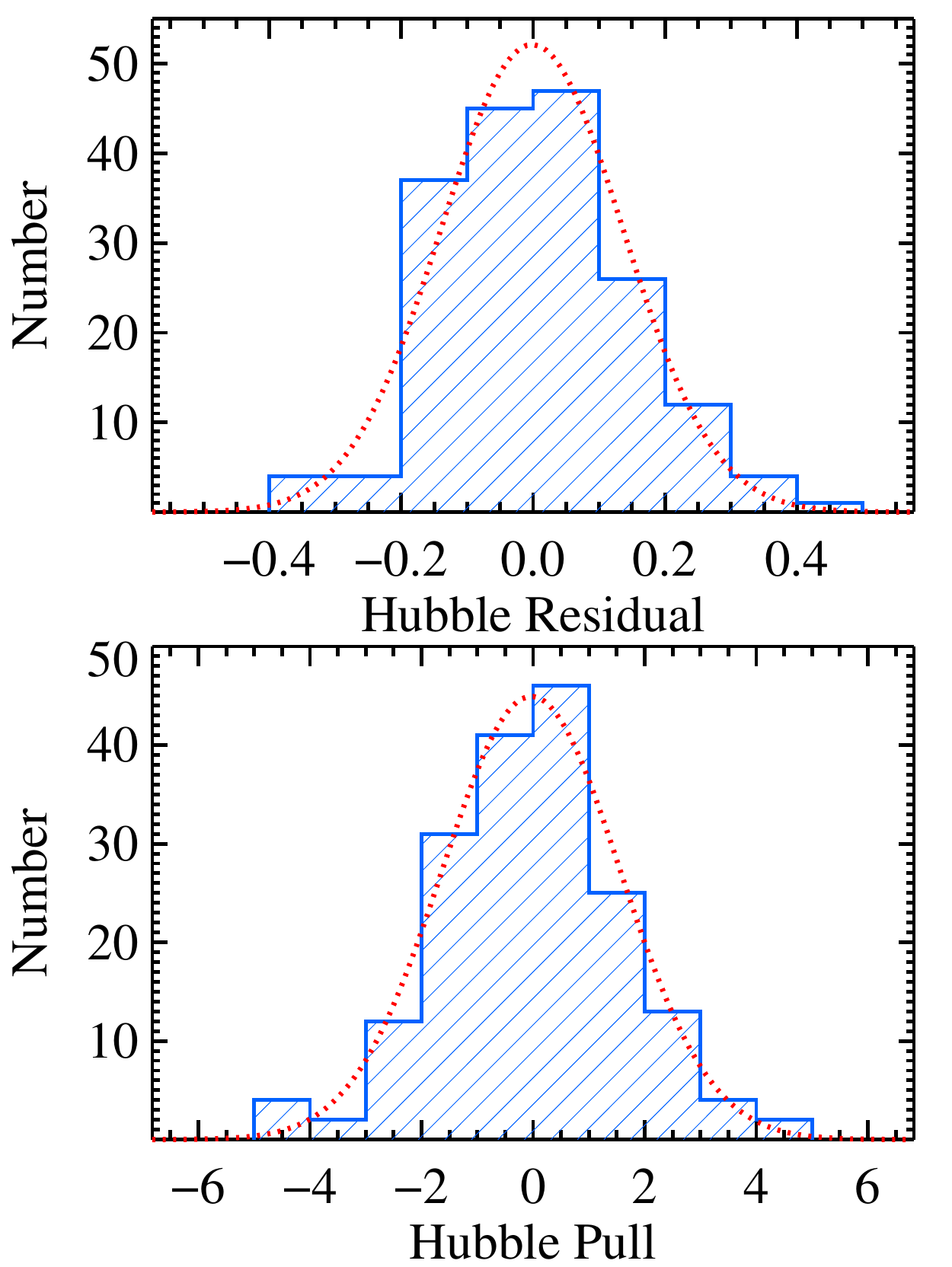}
\caption{Distributions of Hubble residuals (top panel) and pulls
  (bottom panel) for the Foundation DR1 cosmology sample.  The red
  dotted curves represent best-fitting Gaussian functions to the
  distributions.  The best-fitting Gaussian functions have standard
  deviations of 0.138~mag and 1.60 for the Hubble residuals and pulls,
  respectively.}\label{f:res_hist}
\end{center}
\end{figure}

None the less, the Foundation sample Hubble residuals relative to a
fiducial $\Lambda$CDM model are encouraging.  Fitting a Gaussian
function to the residuals, we find a standard deviation of only
0.138~mag.  We also measure a weighted root-mean square (RMS) of
0.136~mag for the Hubble residuals, consistent with the simple
Gaussian measurement.  Similarly fitting a Gaussian function to the
pulls (the residual divided by the uncertainty), we find a standard
deviation of 1.60, close to the expected value of 1 for a normal
distribution with uncertainties exactly matching the full scatter.
Since the standard deviation of the pull is larger than 1, there must
be an additional term, such as intrinsic scatter, that is currently
not included in the uncertainties.  We plot these distributions in
Figure~\ref{f:res_hist}.

In addition to the relatively small Hubble residuals, we find an
intrinsic scatter of only $\sigma_{\rm int} = \sigint$~mag, when we
require that $\chi^{2}/\dof = 1$.  A smaller value of $\sigma_{\rm
  int} = \sigchi$~mag is found when requiring that the standard
deviation of the pulls be 1.  Further improvements to our data
reduction pipeline, the PS1 calibration, the available extant data,
and the distance-fitting algorithm/model should all improve the
intrinsic scatter.

Despite the current success, there are still several aspects that
require additional future scrutiny.  In particular, there is a
moderate correlation between the absolute Hubble residuals and the
redshift uncertainty.  SNe with precise redshifts ($dz < 10^{-3}$)
have median absolute Hubble residuals that are significantly smaller
than of SNe with less precise redshifts ($dz \ge 10^{-3}$): 0.086 and
0.119~mag, respectively.  We expect that more precise redshifts will
improve the distance residuals for the latter group of SNe and will
likely decrease the measured intrinsic scatter and RMS of our sample.

Finally, we note that eight of the nine SNe~Ia that do not pass
Chauvenet's criterion have Hubble residuals of $d\mu \approx
-0.5$~mag.  Six of these objects (SNe~2016ai, 2016aqa, 2016aqt,
2016ayf, 2016hjk, and CSS151120:044526-190158) have spectra that are
possibly consistent with SN~2006gz and other high-luminosity, peculiar
SNe~Ia.  Additionally, these SNe have colour-corrected absolute
magnitudes ($M_{B} = m_{B} - \beta c - \mu$) of $-19.81$ to
$-20.20$~mag with a median of $-20.06$~mag, while the median for the
cosmological subsample is $-19.48$~mag.  Although it is unclear if
these objects are truly similar to SN~2006gz, this is a strong
possibility.  If that were the case, then up to 6.2\% of the
Foundation Supernova sample is comprised of SN~2006gz-like objects, a
surprisingly large number.  The large fraction of SN~2006gz-like
objects in our sample is likely the result of selecting more galaxies
from low-luminosity galaxies, which have a higher relative rate of
this class, compared to previous samples \citep{Khan11:sc,
  Taubenberger11}.  Detailed light-curve and spectral analyses should
improve our classification for these objects.


\section{Discussion and Conclusions}\label{s:disc}

We have presented the survey strategy/design and first results from
the Foundation Supernova Survey.  Over the next several years, we will
use the PS1 telescope to observe hundreds of SNe~Ia at $z < 0.1$.
These well-measured light curves will provide precise and accurate
distances to these SNe, which in turn will provide a foundation for
future SN cosmology analyses.

We have motivated our strategy, stressing that the PS1 system is ideal
for this work.  In particular, PS1 has already observed all positions
north of $-30$ declination, making late-time template observations
unnecessary.  The PS1 system is a well-calibrated photometric system
(to the mmag level) with a precisely and accurately measured
instrument response.  We have already observed \about 500
spectroscopically confirmed and \about 3000 photometrically classified
high-$z$ SNe~Ia (\citealt{Rest14}, Scolnic et al., in prep., Jones et
al., in prep.) with PS1.  All PS1 SNe (both high-$z$ and Foundation)
will be reduced with a single, well-tested data-reduction pipeline.

Our follow-up strategy is economical, yet still provides excellent
light-curve coverage.  At our current rate, we can observe \about 140
SNe~Ia per year, but we believe that we can increase the rate to 200
SNe~Ia per year given additional spectroscopic resources and/or timely
publicly announced classifications.

We already have obtained \nsn\ complete SN~Ia light curves, which we
present here (and release publicly).  From those data, we have derived
light-curve parameters and distance estimates.  We created a Hubble
diagram for the Foundation cosmology sample of \ncosmo\ SNe~Ia,
finding that we have already created a competitive sample that is both
larger than every other homogeneous sample of low-$z$ SNe~Ia and with
low intrinsic scatter ($\sigma_{\rm int} = \sigint$~mag).  This sample
is already comparable in size to the entire current low-$z$ SN~Ia
sample used for cosmological analyses.

Our current sample comes primarily from untargeted SN searches (mostly
ASASSN and PSST), and a large fraction of SNe~Ia have relatively faint
host galaxies.  One-third of all host galaxies did not have a
catalogued redshift, and we provide redshifts here for those galaxies.
There are a surprising number of SNe~Ia that have Hubble residuals of
$d\mu \approx -0.6$~mag.  While these SNe are removed from the
cosmologically useful subsample through various cuts (and in
particular Chauvenet's criterion), these may be peculiar SNe~Ia that
do not have adequate spectroscopic data to distinguish them from more
typical SNe~Ia.

In the coming months and years, we will make further improvements to
our data reduction pipeline, add host-galaxy and spectral data, and
eventually constrain cosmological parameters.  We expect our next data
release to be larger than the entire current low-$z$ sample.

The Foundation Supernova Survey is critical for the success of current
and future surveys such as DES, LSST, and {\it WFIRST}.  As SN
cosmology analyses are currently systematics limited and the largest
systematic is currently the low-$z$ sample, the Foundation Supernova
Survey will have a larger impact on measuring cosmological parameters
than the current generation of high-$z$ surveys.

\section*{Acknowledgements}

{\it Facility:} Pan-STARRS-1 (GPC1); SOAR (Goodman spectrograph);
Mayall (KOSMOS spectrograph); FLWO:1.5m (FAST); ARC (DIS); SALT (RSS);
Keck:I (LRIS); Keck:II (DEIMOS)

\bigskip

We thank the SN community, both professional and amateur, for its
tireless efforts discovering and classifying SNe.  We especially thank
those who make discoveries, classifications, and data immediately
public.  Timely, public access to this information was both critical
for this work and a significant benefit for all of science.  We
especially thank A.\ Castro-Tirado, R.\ Chornock, A.\ Gal-Yam, B.\
Shappee, I.\ Shivvers, J.\ Strader, and O.\ Yaron for providing
additional information about their observations.  We thank S.\
Downing, A.\ Duarte, M.\ Garcia, R.\ Hounsell, A.\ Kniazev, M.\ Kotze,
B.\ Miszalski, B.\ Patel, E.\ Romero Colmenero, R.\ Skelton, and J.\
Tonry for assisting in some of the observations presented here.  We
thank C.\ Stubbs for his continued support of this project and other
intellectual contributions.

This manuscript is based upon work supported by the National
Aeronautics and Space Administration under Contract No.\ NNG16PJ34C
issued through the {\it WFIRST} Science Investigation Teams Programme.
R.J.F.\ and D.S.\ were supported in part by NASA grant 14-WPS14-0048.
The UCSC group is supported in part by NSF grant AST-1518052, the
Gordon \& Betty Moore Foundation, and from fellowships from the Alfred
P.\ Sloan Foundation and the David and Lucile Packard Foundation to
R.J.F.
D.S.\ acknowledges support from NASA through Hubble Fellowship grant
HST-HF2-51383.001 awarded by the Space Telescope Science Institute,
which is operated by the Association of Universities for Research in
Astronomy, Inc., for NASA, under contract NAS 5-26555 and the Kavli
Institute for Cosmological Physics at the University of Chicago
through grant NSF PHY-1125897 and an endowment from the Kavli
Foundation and its founder Fred Kavli.
The research leading to these results has received funding from the
European Research Council under the European Union's Seventh Framework
Programme (FP7/2007-2013)/ERC Grant agreement n$^{\rm o}$ [291222]
(PI: S.J.S.).

Pan-STARRS is supported in part by the National Aeronautics and Space
Administration under Grants NNX12AT65G and NNX14AM74G.  The
Pan-STARRS1 Surveys (PS1) and the PS1 public science archive have been
made possible through contributions by the Institute for Astronomy,
the University of Hawaii, the Pan-STARRS Project Office, the
Max-Planck Society and its participating institutes, the Max Planck
Institute for Astronomy, Heidelberg and the Max Planck Institute for
Extraterrestrial Physics, Garching, The Johns Hopkins University,
Durham University, the University of Edinburgh, the Queen's University
Belfast, the Harvard-Smithsonian Center for Astrophysics, the Las
Cumbres Observatory Global Telescope Network Incorporated, the
National Central University of Taiwan, the Space Telescope Science
Institute, the National Aeronautics and Space Administration under
Grant No.\ NNX08AR22G issued through the Planetary Science Division of
the NASA Science Mission Directorate, the National Science Foundation
Grant No.\ AST--1238877, the University of Maryland, Eotvos Lorand
University (ELTE), the Los Alamos National Laboratory, and the Gordon
and Betty Moore Foundation.

This paper is based on observations (NOAO Prop.\ IDs: 2015A-0253 and
2015B-0313; PI: R.J.F.) obtained at the Southern Astrophysical
Research (SOAR) telescope, which is a joint project of the
Minist\'erio da Ci\^{e}ncia, Tecnologia, e Inova\c{c}\~ao (MCTI) da
Rep\'{u}blica Federativa do Brasil; the U.S. National Optical
Astronomy Observatory (NOAO); the University of North Carolina at
Chapel Hill (UNC); and Michigan State University (MSU), and at Kitt
Peak National Observatory, NOAO, which is operated by the Association
of Universities for Research in Astronomy (AURA) under cooperative
agreement with the National Science Foundation. The authors are
honored to be permitted to conduct astronomical research on Iolkam
Du'ag (Kitt Peak), a mountain with particular significance to the
Tohono O'odham.

Some of the data presented herein were obtained at the W.\ M.\ Keck
Observatory, which is operated as a scientific partnership among the
California Institute of Technology, the University of California and
the National Aeronautics and Space Administration. The Observatory was
made possible by the generous financial support of the W.\ M.\ Keck
Foundation.  The authors wish to recognize and acknowledge the very
significant cultural role and reverence that the summit of Maunakea
has always had within the indigenous Hawaiian community.  We are most
fortunate to have the opportunity to conduct observations from this
mountain.

Research at Lick Observatory is partially supported by a generous
gift from Google.

Some of the observations reported in this paper were obtained with the
Southern African Large Telescope (SALT), and we thank the SALT
Astronomers for assistance.

Based on observations obtained with the Apache Point Observatory
3.5-meter telescope, which is owned and operated by the Astrophysical
Research Consortium.

We acknowledge use of some data collected at the European Organisation
for Astronomical Research in the Southern Hemisphere, Chile as part of
PESSTO, (the Public ESO Spectroscopic Survey for Transient Objects
Survey) ESO programme 188.D-3003, 191.D-0935.

This manuscript was initiated during the ``Dynamic Universe:
Understanding ExaScale Astronomical Synoptic Surveys'' workshop at the
Aspen Center for Physics, which is supported in part by the NSF under
grant No.\ PHYS-1066293.  R.J.F.\ and D.S.\ thank the Aspen Center for
Physics for its hospitality during the ``Dynamic Universe'' workshop
in June 2015.


\bibliographystyle{mnras}
\bibliography{../astro_refs}

\onecolumn
\begin{landscape}
\begin{deluxetable}{lllllll}
\tabletypesize{\footnotesize}
\tablewidth{0pt}
\tablecaption{Foundation Supernova Sample Data\label{t:sample}}
\tablehead{
\colhead{IAU Name} & \colhead{Discovery Name} & \colhead{Alternative Names} & \colhead{Galaxy} & \colhead{Discovery Ref.} & \colhead{Classification Ref.} & \colhead{Type}}

\startdata

2015I & PSN J07174570+2320406           &                                   & NGC 2357                     & CBET 4106 & ATEL 7476 & \\
\cline{1-7}
\multirow{3}{*}{2015M} & \multirow{3}{*}{PSN J13003230+2758411}           & CSS150521:130032+275841 & \multirow{3}{*}{COMAi J130032.301+275841.02} & \multirow{2}{*}{CBET 4123} & ATEL 7528 & \multirow{3}{*}{06gz}\\
 & & KISS15n & &  \multirow{2}{*}{ATEL 7532} & ATEL 7529 & \\
 & & PS15apc & & & ATEL 7532 & \\
\cline{1-7}
2015ar  & PSN J01072038+3223598         & CSS151130:010720+322359           & NGC 383                      & CBET 4193 & ATEL 8291 & \\
\cline{1-7}
2015az  & PSN J03094251+4058293         & PS15dqn                           & IC 290                       & CBET 4208 & ATEL 8348 & \\
\cline{1-7}
2016E   & PS16n                         &                                   & 2MASX J05034387+0324429      & ATEL 8524 & ATEL 8524 & \\
\cline{1-7}
2016F   & ASASSN-16ad                   & CSS160204:013932+334936           & KUG 0136+335                 & ATEL 8521 & ATEL 8532       & \\
\enddata

\vspace{-0.7cm}

\tablecomments{S16 = \citet{Smartt16}}

\tablenotetext{a}{SN~2017po is hosted by a galaxy pair.  It is
  physically closer to the smaller galaxy, CGCG 223-033 NED01, but we
  consider it to be more likely to be hosted by the larger galaxy,
  UGC~10166.}
\end{deluxetable}
\end{landscape}
\twocolumn
\begin{deluxetable}{lcrrrr}
\tablewidth{0pt}
\tabletypesize{\footnotesize}
\tablecaption{PS1 Photometry \label{t:lc}}
\tablecolumns{6}
\tablehead{
\colhead{MJD} & \colhead{Filter} & \colhead{Magnitude\tablenotemark{1}} & \colhead{Uncertainty} & \colhead{Flux\tablenotemark{2}} & \colhead{Uncertainty}}
\startdata
 
\hline
\multicolumn{6}{c}{2015I}\\
\hline
57147.250 & $g_{\rm P1}$ & 14.999 & 0.004 & 100070.0 &    398.2\\
57153.254 & $g_{\rm P1}$ & 14.180 & 0.004 & 212910.0 &    713.3\\
57160.258 & $g_{\rm P1}$ & 14.126 & 0.004 & 223670.0 &    751.7\\
57167.258 & $g_{\rm P1}$ & 14.502 & 0.005 & 158270.0 &    700.7\\
57147.250 & $r_{\rm P1}$ & 14.953 & 0.004 & 104410.0 &    351.0\\
57153.254 & $r_{\rm P1}$ & 14.190 & 0.003 & 210800.0 &    624.9\\
57160.258 & $r_{\rm P1}$ & 14.056 & 0.003 & 238630.0 &    642.7\\
\enddata
 
\tablenotetext{1}{Limiting magnitudes are 3~$\sigma$.}
\tablenotetext{2}{Flux units are such that $m = 27.5 - 2.5 \log_{10} (f)$.}
 
\end{deluxetable}
\onecolumn

\begin{deluxetable}{llllcc}
\tabletypesize{\footnotesize}
\tablewidth{0pt}
\tablecaption{Uncatalogued Redshifts\label{t:redshift}}
\tablehead{
\colhead{SN} & \colhead{Galaxy} & \colhead{Redshift} & \colhead{Redshift}    & \colhead{Derived} & \colhead{Telescope}\\
\colhead{}   & \colhead{}       & \colhead{        } & \colhead{Uncertainty} & \colhead{From}    & \colhead{/ATEL\tablenotemark{1}}}

\startdata
2016E                   & 2MASX J05034387+0324429      & 0.05273  & 0.00002  & nucleus     & SOAR\\
2016J                   & Anonymous                    & 0.02642  & 0.00006  & nucleus     & KPNO\\
2016afb                 & Anonymous                    & 0.0677   & 0.0010   & SN position & KPNO\\
2016ajl                 & 2MASX J04422451-2143312      & 0.06741  & 0.00005  & nucleus     & SOAR\\
2016alt                 & GALEXASC J063701.16+441611.9 & 0.090    & 0.006    & SN          & UH 2.2-m\\
2016aqb                 & 2MASXi J1121579-135849       & 0.06289  & 0.00004  & nucleus     & KPNO\\
2016aqs                 & Anonymous                    & 0.02507  & 0.00003  & nucleus     & KPNO\\
2016asf                 & KUG 0647+311                 & 0.018019 & 0.000015 & nucleus     & KPNO\\
\enddata

\tablenotetext{1}{If just a telescope is listed, the spectrum was
  obtained by us and (re-)analysed for this study.  If just an ATEL is
  listed, we simply list the redshift reported in that ATEL.  If both
  a telescope and ATEL are listed, we analyzed the publicly available
  spectrum, which was either submitted to TNS or WISERep, and report
  our derived redshift.}


\end{deluxetable}

\begin{deluxetable}{lllll}
\tabletypesize{\footnotesize}
\tablewidth{0pt}
\tablecaption{Significant Redshift Differences\label{t:zdiff}}
\tablehead{
\colhead{SN} & \colhead{Reported $z$} & \colhead{Updated $z$} & \colhead{Absolute $z$ Difference} & \colhead{$\mu$ Bias (mag)}}

\startdata

2016J                    & 0.034                   & 0.0264217               & 0.0075783 & 0.560 \\
2016aqs                  & 0.03                    & 0.0250655               & 0.0049345 & 0.398 \\
2016asf                  & 0.021                   & 0.0180194               & 0.0029806 & 0.337 \\
2016blg                  & 0.07                    & 0.0610813               & 0.0089187 & 0.310 \\
2016cck                  & 0.041\tablenotemark{1}  & 0.03557                 & 0.00543   & 0.317 \\
2016cvv                  & 0.038                   & 0.0445710               & 0.006571  & $-0.357$ \\
2016eee                  & 0.026                   & 0.03029                 & 0.00429   & $-0.339$ \\
2016gfr                  & 0.014                   & 0.016693                & 0.002693  & $-0.386$ \\
\enddata

\tablenotetext{1}{\citet{Piascik16} initially reported a redshift of
  0.04, but \citet{Zhang16:16cck} reported a redshift of 0.037, which
  is only moderately different from our measured value of 0.03557.}

\tablenotetext{2}{The SN is most similar to SN~2000cx
  \citep{Li01:00cx} at $z = 0.060 \pm 0.002$.  However, it is also
  somewhat similar to SN~1991bg-like objects at $z = 0.047 \pm 0.007$.}

\tablenotetext{3}{\citet{Fraser17} initially reported a redshift of
  0.04, but \citet{Morrell17} reported a NED redshift of 0.033253,
  which is only slightly different from our preferred value of
  0.0332378.}

\end{deluxetable}

\begin{deluxetable}{lcc}
\tabletypesize{\footnotesize}
\tablewidth{0pt}
\tablecaption{Classification Differences\label{t:cdiff}}
\tablehead{
\colhead{SN} & \colhead{Reported Class.} & \colhead{Updated Class.}}

\startdata

2015M                 & SN~1991T-like  & SN~2006gz-like\\
2016aex               & SN~1991T-like? & SN~1991T-like\\
2016alt               & Normal         & SN~2006gz-like\\
2016aqa               & Normal         & SN~1991T-like\\
2016aqb               & Normal         & SN~1991T-like\\
2016arc               & Normal         & SN~1991bg-like\\
2016ayf               & Normal         & SN~1991T-like\\
2016enx               & SN~1991T-like? & SN~1991T-like\\
2017jl                & Normal         & SN~1991T-like?\\
2017lc                & Normal         & SN~2000cx-like?; SN 1991bg-like?\\
2017mu                & Normal         & SN~2006gz-like\\
PS15zn                & SN~Ia          & SN~Ic\\
PS15bwh               & Normal         & SN~1991T-like\\
PS15cwx               & SN~1991T-like  & Normal\tablenotemark{1}\\
PSN J23102264+0735202 & Normal         & SN~2006gz-like\tablenotemark{2}

\enddata

\tablenotetext{1}{Spectrum provided by C.\ Kilpatrick, private comm.}

\tablenotetext{2}{\citet{Elias-Rosa15} remark that at the nominal host
  redshift that the SN would have a low \ion{Si}{II} ejecta velocity,
  and suggest that the host galaxy was misidentified.  However, the
  spectrum is more consistent with SN~2006gz at the nominal redshift
  (see also Table~\ref{t:zdiff}).}

\end{deluxetable}

\begin{deluxetable}{lrrrrrr}
\tablewidth{0pt}
\tabletypesize{\footnotesize}
\tablecaption{Foundation Sample Light-curve Parameters\label{t:param}}
\tablecolumns{7}
\tablehead{
\colhead{SN} & \colhead{$z_{\rm helio}$} & \colhead{$z_{\rm CMB}$} & \colhead{Peak MJD} & \colhead{$x_{1}$} & \colhead{$c$} & \colhead{$m_{B}$ (mag)}}
\startdata
 
2015I                         & 0.00759 (0.00001) & 0.00875 (0.00001) & 57157.10 (0.14) & $-0.20$ (0.19) & $ 0.035$ (0.032) & 13.67 (0.05)\\
2015M                         & 0.02316 (0.00017) & 0.02281 (0.00017) & 57168.67 (0.13) & $ 0.42$ (0.08) & $ 0.005$ (0.027) & 15.24 (0.03)\\
2015ar                        & 0.01700 (0.00004) & 0.01635 (0.00004) & 57352.05 (0.17) & $-1.96$ (0.19) & $-0.092$ (0.038) & 14.77 (0.06)\\
2015az                        & 0.02012 (0.00010) & 0.02133 (0.00010) & 57362.09 (0.15) & $ 0.13$ (0.08) & $ 0.018$ (0.036) & 14.90 (0.10)\\
2016E                         & 0.05273 (0.00002) & 0.05318 (0.00002) & 57392.44 (0.21) & $-0.09$ (0.17) & $-0.055$ (0.035) & 17.41 (0.06)\\
2016F                         & 0.01613 (0.00001) & 0.01513 (0.00001) & 57406.45 (0.09) & $ 0.21$ (0.07) & $ 0.010$ (0.028) & 14.88 (0.04)\\
2016H                         & 0.04611 (0.00001) & 0.04748 (0.00001) & 57400.28 (0.21) & $-0.63$ (0.11) & $ 0.003$ (0.028) & 17.42 (0.04)\\
2016J                         & 0.02642 (0.00006) & 0.02720 (0.00006) & 57400.64 (0.16) & $ 0.54$ (0.13) & $ 0.067$ (0.041) & 16.05 (0.14)\\
\enddata
 
\end{deluxetable}

\begin{deluxetable}{lrrr}
\tabletypesize{\footnotesize}
\tablewidth{0pt}
\tablecaption{Sample Cuts\label{t:cuts}}
\tablehead{
\colhead{Criterion} & \colhead{No.\ of SNe} & \colhead{Cum.\ No.\ of} & \colhead{SNe}\\
\colhead{} & \colhead{Not Passing} & \colhead{SNe Not Passing} & \colhead{Remaining}}

\startdata
Initial Sample                  & \nodata & \nodata & 225 \\
Not Similar to SNe~Iax,         & \multirow{2}{*}{12} & \multirow{2}{*}{12} & \multirow{2}{*}{213} \\
\hspace{1em} SN~1991bg, or SN~2006gz    &    &    &     \\
$E(B-V)_{\rm MW} < 0.25$~mag    &  5 & 17 & 208 \\
Sufficient LC coverage          &  3 & 20 & 205 \\
$t_{\rm first} < 7$~days        &  7 & 25 & 200 \\
$\sigma(x_{1}) < 1$             &  1 & 25 & 200 \\
$\sigma(t_{\rm peak}) < 1$~days &  7 & 31 & 194 \\
$-0.3 < c < 0.3$                & 11 & 36 & 189 \\
$-3.0 < x_{1} < 3.0$            &  2 & 36 & 189 \\
Chauvenet's criterion           &  9 & 45 & 180

\enddata

\vspace{-0.7cm}

\end{deluxetable}

\twocolumn


\end{document}